\documentclass[a4paper,12pt]{article}
\usepackage{amsmath}
\usepackage{amssymb}
\usepackage{graphicx}
\usepackage{color}
\usepackage[normalem]{ulem}

\usepackage[T1]{fontenc}
\usepackage{mathrsfs}             
\usepackage{slashed}              
\usepackage{xspace}
\usepackage{placeins} 
\usepackage{xcolor}
\usepackage{caption}
\usepackage{subcaption}

\makeatletter
\g@addto@macro\bfseries{\boldmath}
\makeatother

\evensidemargin \oddsidemargin
\numberwithin{equation}{section}
\numberwithin{table}{section}
\numberwithin{figure}{section}

\def \atlas {ATLAS\xspace}
\def \babar {BaBar\xspace}
\def \belleone {Belle\xspace}
\def \belletwo {Belle\,II\xspace}

\def \codexb {\hbox{CODEX-b}\xspace}
\def \cresstiii {CRESST-III\xspace}

\def \faser {FASER\xspace}

\def \lhcb {LHCb\xspace}
\def \lthree {L3\xspace}

\def \microboone {MicroBooNE\xspace}

\def \scalar {\ensuremath{\phi}\xspace}
\def \mscalar {\ensuremath{m_{\scalar}}\xspace}
\def \pot {protons on target\xspace}
\def \kaon {kaon\xspace}
\def \pion {pion\xspace}
\def \kaons {kaons\xspace}
\def \pions {pions\xspace}
\def \klong {\ensuremath{K^0_L}\xspace}
\def \kstarz {\ensuremath{K^*(892)^0}\xspace}
\def \pbinv {\ensuremath{\,\mathrm{pb}^{-1}}\xspace}
\def \fbinv {\ensuremath{\,\mathrm{fb}^{-1}}\xspace}
\def \abinv {\ensuremath{\,\mathrm{ab}^{-1}}\xspace}
\def \bmeson {\ensuremath{B\text{-meson}}\xspace}
\def \bmesons {\ensuremath{B\text{-mesons}}\xspace}

\def \pizero {\ensuremath{\pi^0}\xspace}
\def \bf {\ensuremath{\mathcal{B}}\xspace}
\def \sqrts {\ensuremath{\sqrt{s}}\xspace}

\newcommand{\mm}{\ensuremath{\mathrm{\,mm}}\xspace}
\newcommand{\cm}{\ensuremath{\mathrm{\,cm}}\xspace}
\newcommand{\m}{\ensuremath{\mathrm{\,m}}\xspace}
\newcommand{\msq}{\ensuremath{\mathrm{\,m^2}}\xspace}
\newcommand{\gev}{\ensuremath{\mathrm{\,Ge\kern -0.1em V}}\xspace}
\newcommand{\mev}{\ensuremath{\mathrm{\,Me\kern -0.1em V}}\xspace}
\newcommand{\tev}{\ensuremath{\mathrm{\,Te\kern -0.1em V}}\xspace}

\usepackage[colorlinks,citecolor=blue,linktoc=all,linkcolor=teal]{hyperref}

\begin{document}

\begin{flushright}
P3H-23-034, TTP23-018
\end{flushright} 

\title{\vspace{0.2cm} \textbf{Dark Higgs Bosons at Colliders}\vspace{0.8cm}}

\author{Torben Ferber$^1$, Alexander Grohsjean$^2$,  and Felix Kahlhoefer$^3$ 
\\
\\
\small${}^{1}$Institute of Experimental Particle Physics (ETP), \\\small Karlsruhe Institute of Technology (KIT), 76131 Karlsruhe, Germany \\[0.2cm]
\small${}^2$Institute of Experimental Particle Physics,\\\small University of Hamburg, 22761 Hamburg, Germany \\[0.2cm]
\small${}^3$Institute for Theoretical Particle Physics (TTP),\\\small Karlsruhe Institute of Technology (KIT), 76131 Karlsruhe, Germany\\[0.2cm]
\small E-mail: \href{mailto:torben.ferber@kit.edu}{torben.ferber@kit.edu}, \href{mailto:alexander.grohsjean@desy.de}{alexander.grohsjean@desy.de}, \href{mailto:kahlhoefer@kit.edu}{kahlhoefer@kit.edu}}
\date{}

{\let\newpage\relax\maketitle}

\vspace{0.2cm}

\begin{abstract}
The Large Hadron Collider (LHC) has confirmed the Higgs mechanism to generate mass in the Standard Model (SM), making it attractive also to consider spontaneous symmetry breaking as the origin of mass for new particles in a dark sector extension of the SM. Such a dark Higgs mechanism may in particular give mass to a dark matter candidate and to the gauge boson mediating its interactions (called dark photon). In this review, we summarize the phenomenology of the resulting dark Higgs boson and discuss the corresponding search strategies with a focus on collider experiments.  We consider both the case that the dark Higgs boson is heavier than the SM Higgs boson, in which case leading constraints come from direct searches for new Higgs bosons as well missing-energy searches at the LHC, and the case that the dark Higgs boson is (potentially much) lighter than the SM Higgs boson, such that the maximum sensitivity comes from electron-positron colliders and fixed-target experiments. Of particular experimental interest for both cases is the associated production of a dark Higgs boson with a dark photon, which subsequently decays into SM fermions, dark matter particles or long-lived dark sector states. We also discuss the important role of exotic decays of the SM-like Higgs boson and complementary constraints arising from early-universe cosmology, astrophysics, and direct searches for dark matter in laboratory experiments. \\[0.2cm]
\end{abstract}

\newpage

\tableofcontents

\newpage

\section{Introduction}
\label{sec:intro}
Following the discovery of a Standard Model (SM)-like Higgs boson at the Large Hadron Collider (LHC), the ATLAS and CMS collaborations have been able to map out in detail its properties and find consistency with the assumption that it couples to other SM particles proportionally to their mass~\cite{ATLAS:2022vkf,CMS:2022dwd}. These results demonstrate in a compelling way that the Higgs mechanism is responsible for generating the masses of elementary particles.
Nevertheless, the origin of the majority of mass in the universe remains a mystery, because it is not in the form of known particles but in a completely new form called dark matter (DM). If DM is also composed of elementary particles, we can hope to detect these particles in the laboratory or infer their properties from astrophysical observations. Many models predicting observable signatures have been proposed and a worldwide effort to carry out the corresponding measurements is underway, see Ref.~\cite{Arbey:2021gdg} for a recent review.

A particularly intriguing possibility is that DM particles obtain at least part of their mass in a way that is reminiscent of electroweak symmetry breaking, i.e.\ by coupling to a complex scalar field, which obtains a vacuum expectation value (vev) that spontaneously breaks a new gauge symmetry. This mechanism is referred to as the \emph{dark Higgs mechanism}, and the resulting scalar boson is called the dark Higgs boson.\footnote{To the best of our knowledge, the term was first proposed in Ref.~\cite{Weihs:2011wp}, although very similar models had already been proposed earlier, see e.g.\ Refs.~\cite{Schabinger:2005ei,Patt:2006fw,Batell:2009yf}. In many regards, the model that we consider is identical to the Hidden Abelian Higgs Model (HAHM) proposed in Ref.~\cite{Wells:2008xg}.} The key implication of this idea is that DM particles do not arise in isolation, but together with additional scalar and vector bosons, forming an entire \emph{dark sector}.

The presence of a dark Higgs boson and a new gauge symmetry beyond the SM provides exciting new possibilities to study the phenomenology of DM. Not only do the additional states and interactions offer new ways for DM particles to be produced in the laboratory, but they themselves become an object worthy of study. Indeed, even if DM particles are too heavy or too weakly coupled to be found in the laboratory, we may be able to produce and detect dark Higgs bosons. This is because scalar fields generally mix with each other through the scalar potential. The dark Higgs boson hence inherits some properties of its SM counterpart, and in turn, modifies the experimental signatures of the latter. In other words, the discovery of the SM-like Higgs boson at the LHC may pave the path also for the discovery of dark Higgs bosons.

In this review, we discuss the properties of dark Higgs bosons and present an overview of existing experimental results. We start with the simplest case, in which the dark Higgs boson is the only accessible state beyond the SM and its couplings are induced by Higgs mixing, and gradually add additional production and decay modes, as well as additional states that can be produced together with dark Higgs bosons and give rise to richer experimental signatures. While our primary focus is on collider and accelerator experiments, we also discuss the resulting DM phenomenology and its implications for astrophysics and cosmology. 

\newpage
\section{Terminology and outline}

A dark Higgs field is a complex scalar field, which is a singlet under the SM gauge group but carries charge under a new $U(1)'$ gauge group.\footnote{Constructions with non-Abelian gauge groups also exist in the literature~\cite{Zhang:2009dd,Baek:2013dwa,Ko:2016fcd}, but we will focus on the simpler Abelian case here.} 
The corresponding Lagrangian is given by
\begin{equation}
 \mathcal{L}_\Phi = \left[ \left( \partial^\mu + i g^\prime q_\Phi A^{\prime \mu} \right) \Phi\right]^\dagger\left[ \left( \partial_\mu + i g^\prime q_\Phi A^\prime_\mu \right) \Phi\right] - V(\Phi, H) \; ,
\end{equation}
where $\Phi$ ($H$) denotes the complex dark Higgs (SM Higgs) field, $A'$ denotes the $U(1)'$ gauge boson, and $g^\prime$ and $q_\Phi$ are the $U(1)'$ gauge coupling and the charge of the dark Higgs boson, respectively. The dark Higgs field acquires a vacuum expectation value, thereby breaking the $U(1)'$ gauge symmetry spontaneously and giving mass to the corresponding gauge boson. In the process, it may also give mass to other dark sector particles.
We will denote the vacuum expectation value of the dark Higgs field by $w$ and the resulting physical dark Higgs boson by $\phi$. The gauge boson mass is then given by $m_{A'} = g^\prime q_\Phi w$.

Our definition of a dark Higgs boson has the central implication  that $\phi$ will in general couple linearly to other fields in the theory. In particular, it is generally expected to be unstable and decay into any pair of particles 1 and 2 with $m_1 + m_2 < m_\phi$. If the dark Higgs bosons decay dominantly into stable (i.e. invisible) dark sector particles, it will be challenging (but not impossible) to observe them experimentally. 
In the following, we will therefore focus primarily on the case that the dark Higgs boson is the lightest state in the dark sector. In this case, the only possible decay modes are those involving SM particles. 

Indeed, such decay modes are generally induced via Higgs mixing. Before symmetry breaking the scalar potential contains a term of the form
\begin{equation}
 V(\Phi, H) \supset \lambda_{h\phi} |\Phi|^2 |H|^2 \; .
\end{equation}
After symmetry breaking and the replacements
\begin{align}
\Phi & \to \frac{\phi + w}{\sqrt{2}} \nonumber \\
H & \to \frac{1}{\sqrt{2}} \begin{pmatrix} 0 \\ h + v \end{pmatrix} \nonumber
\end{align}
this term leads to mixing between the SM Higgs boson $h$ and the dark Higgs boson $\phi$ with mixing angle $\theta$ given by\footnote{This expression assumes $\theta \ll 1$, as required by the observed properties of the SM-like Higgs boson. The general expression that is valid also for large mixing can be found e.g.\ in Refs.~\cite{Baek:2011aa,Duerr:2016tmh}.}
\begin{equation}
\theta \approx \frac{\lambda_{h\phi} \, v \, w}{m_h^2 - m_\phi^2} \; .
\end{equation}

The mixing between the two Higgs bosons can be captured by the replacement
\begin{align}
 h & \to \cos \theta \, h + \sin \theta \, \phi \\
 \phi & \to - \sin \theta \, h + \cos \theta \, \phi \; ,
\end{align}
which leads to three main consequences:
\begin{enumerate}
    \item The dark Higgs boson obtains couplings to SM particles proportional to $\sin \theta$. It will therefore have the same decay modes as an SM-like Higgs boson with mass $m_\phi$, with each partial decay width suppressed by a factor $\sin^2 \theta$.
    \item The couplings of the SM-like Higgs boson to the other SM particles are suppressed proportional to $\cos \theta$, which may be in conflict with the  observed agreement of the Higgs signal strength with SM expectations.
    \item The SM-like Higgs boson obtains couplings to dark sector particles proportional to $\sin \theta$, which may induce new decay modes and thereby shift the branching ratios away from the SM prediction.
\end{enumerate}
Furthermore, for $m_\phi < m_h/2$ the SM Higgs boson can decay into a pair of dark Higgs bosons, whereas the opposite decay is possible for $m_h < m_\phi / 2$.

In the simplest case, where no other dark sector states are kinematically accessible, the phenomenology of a dark Higgs boson can be fully characterized by its mass $m_\phi$ and the mixing angle $\sin \theta$. We review the production and decay modes of this simple scenario in section~\ref{sec:pheno}. Constraints from the observed properties of the SM-like Higgs boson and from direct searches for dark Higgs bosons at accelerators are discussed in  sections~\ref{sec:Signal_strength} and~\ref{sec:direct_Higgs}, respectively.

A straightforward extension of this model that has received great interest in recent years allows for additional interactions of the new gauge boson, which is commonly called dark photon (see Refs.~\cite{Fabbrichesi:2020wbt,Caputo:2021eaa} for recent reviews). Such couplings arise either directly if SM fermions carry a charge under the $U(1)'$ gauge group, or indirectly through kinetic mixing. In both cases, the resulting interaction Lagrangian after diagonalization can be written as
\begin{equation}
    \mathcal{L}_{A'} = -\frac{1}{4} F^{\prime \mu \nu} F_{\mu\nu}^\prime - A^{\prime \mu} \sum_{f} g_f \bar{f} \gamma_\mu f \; ,
\end{equation}
where $F^{\prime \mu \nu} = \partial^\mu A^{\prime \nu} - \partial^\nu A^{\prime \mu}$, $f$ denotes the various SM fermions and $g_f$ their effective couplings.
As a result, the dark photon provides a new way to produce dark Higgs bosons via dark Higgs-strahlung: $\text{SM SM} \to A'^\ast \to A' + \phi$, as well as a new decay mode: $\phi/h \to 2 A' \to 4 \, \text{SM}$. This makes searching for dark Higgs bosons possible even if the Higgs mixing angle $\theta$ is tiny. We discuss the corresponding experimental strategies in section~\ref{sec:Higgs_dark_photon}.

The main motivation to study dark Higgs bosons is to address the DM puzzle. To do so, the model introduced above can be easily extended by a third dark sector particle, namely a fermion that is an SM singlet and carries charge under the $U(1)'$ gauge group.\footnote{The case of a complex scalar DM candidate is also possible~\cite{Baek:2014kna} and leads to similar (but potentially distinguishable) phenomenology~\cite{Ko:2016xwd,Kamon:2017yfx}. For the case of vector DM, we refer to Ref.~\cite{Baek:2012se}.} The broken gauge symmetry leads to a discrete symmetry, which ensures the stability of the fermion, such that it becomes an attractive DM candidate. The simplest (anomaly-free) version of this model introduces two chiral fermions $\chi_\mathrm{L,R}$ with opposite charge $\pm q_\text{DM}$ under the $U(1)'$ gauge symmetry. Assuming that the dark Higgs boson carries charge $-2 q_\text{DM}$, one obtains the following gauge-invariant Lagrangian:
\begin{equation}
\mathcal{L}_\chi  = {i \bar{\chi} \slashed{\partial} \chi - y_\chi \bar{\chi^c} \left(P_L \Phi + P_R \Phi^\ast \right) \chi - g_\chi A'^\mu \bar{\chi} \gamma_\mu \gamma^5 \chi } \; , \label{eq:Majorana}
\end{equation}
where $\chi = (\chi_L, \chi^c_R)$, $y_\chi$ denotes the Yukawa coupling between the dark Higgs boson and the DM particle, $g_\chi = g' q_\text{DM}$ denotes the effective coupling to the dark photon and $P_{L,R} = \tfrac{1}{2}(1 \mp \gamma^5)$. We note that this Lagrangian assumes that the DM mass is generated exclusively via the dark Higgs field, such that $m_\chi = y_\chi \, w / \sqrt{2}$. This setup opens up the possibility to search for dark Higgs bosons in events with missing energy, arising from invisible decays of either the dark Higgs boson or the dark photon, as discussed in sections~\ref{sec:invisible_Higgs} and~\ref{sec:Higgs_MET}.

As a final extension of the model one can include a gauge-invariant Dirac mass term $m_\mathrm{D}\bar{\chi}\chi$ in addition to the Majorana mass terms $m_L \bar{\chi^c_L} \chi_L$ and $m_R \bar{\chi^c_R} \chi_R$ generated by the dark Higgs mechanism. This results in the mass matrix
\begin{equation}
 \mathcal{L}_\chi \supset -\frac{1}{2} \begin{pmatrix} \overline{\chi_L^c} & \overline{\chi_R} \end{pmatrix} \begin{pmatrix} m_L & m_D \\ m_D & m_R \end{pmatrix} \begin{pmatrix}
\chi_L \\ \chi_R^c
\end{pmatrix}
+ \text{h.c.} \;,
\end{equation}
which is diagonalized by two (Majorana) mass eigenstates $\psi_1$ and $\psi_2$ that satisfy the relation
\begin{align}
 \chi_L &= \cos \theta \, \psi_{1,L} + i \sin \theta  \, \psi_{2,L} , \\
 \chi_R &= \sin \theta  \, \psi_{1,R} + i \cos \theta  \, \psi_{2,R}
\end{align}
with the mixing angle $\theta$. Assuming $m_L \approx m_R \ll m_D$, the mixing angle is close to $\pi/4$ and diagonalization of the mass matrix then leads to so-called Pseudo-Dirac~\cite{DeSimone:2010tf} (or inelastic~\cite{Tucker-Smith:2001myb}) DM: two Majorana fermions $\chi_1$ and $\chi_2$ with small mass splitting $\Delta = m_2 - m_1 \ll m_1, m_2$ and off-diagonal couplings to the dark photon. This type of model has received great interest~\cite{Davoli:2017swj,Izaguirre:2015zva,Berlin:2018jbm,Duerr:2020muu} due to its ability to evade direct and indirect detection constraints, while at the same time predicting exciting new signatures at accelerators resulting from the production of long-lived $\chi_2$ particles. We will discuss the experimental implications of such a setup in section~\ref{sec:Higgs_DV}.

To conclude our review of dark Higgs bosons, we consider in section~\ref{sec:cosmology_and_astro} complementary constraints from astrophysics and cosmology, as well as from DM direct and indirect detection experiments.

\FloatBarrier
\section{Phenomenology of dark Higgs bosons}
\label{sec:pheno}
\begin{figure}[t]%
\centering
     \begin{subfigure}[b]{0.25\textwidth}
         \centering
         \includegraphics[width=\textwidth]{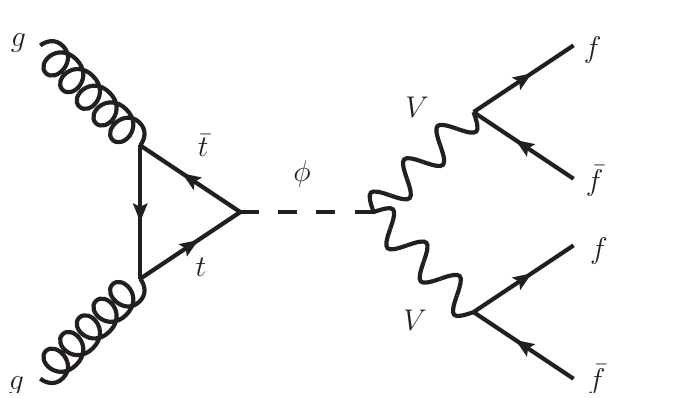}
         \caption{Gluon fusion production.}\label{subfig:production_feynman:gg_4f}
     \end{subfigure}
     \hspace{0.075\textwidth}
     \begin{subfigure}[b]{0.25\textwidth}
         \centering
         \includegraphics[width=\textwidth]{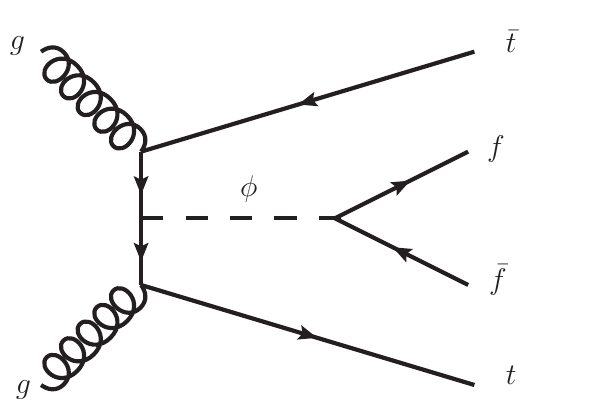}
         \caption{Top-associated production.}\label{subfig:production_feynman:gg_darkscalar_toptop}
     \end{subfigure}
     \hspace{0.075\textwidth}
     \begin{subfigure}[b]{0.25\textwidth}
         \centering
         \includegraphics[width=\textwidth]{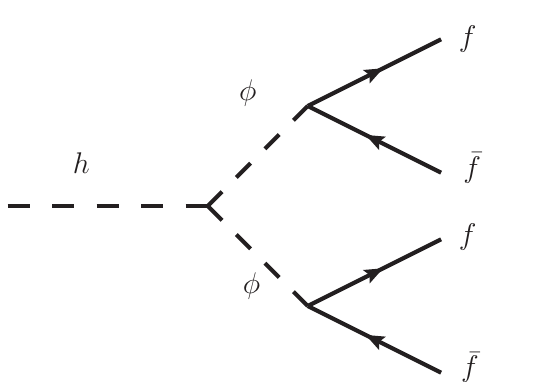}
         \caption{Decay of the SM-like Higgs boson.}
         \label{subfig:production_feynman:higgs_dark_scalar_dark_scalar}
     \end{subfigure}
     
     \begin{subfigure}[b]{0.32\textwidth}
         \centering
         \includegraphics[width=\textwidth]{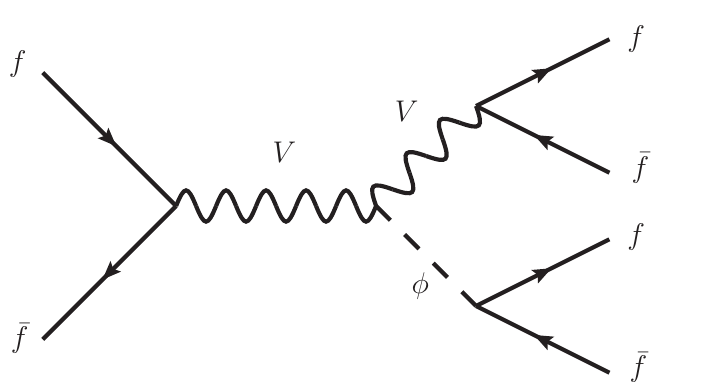}
         \caption{Dark Higgs-strahlung.\\ \quad}
         \label{subfig:production_feynman:darkphoton}
     \end{subfigure}
     \begin{subfigure}[b]{0.32\textwidth}
         \centering
         \includegraphics[width=\textwidth]{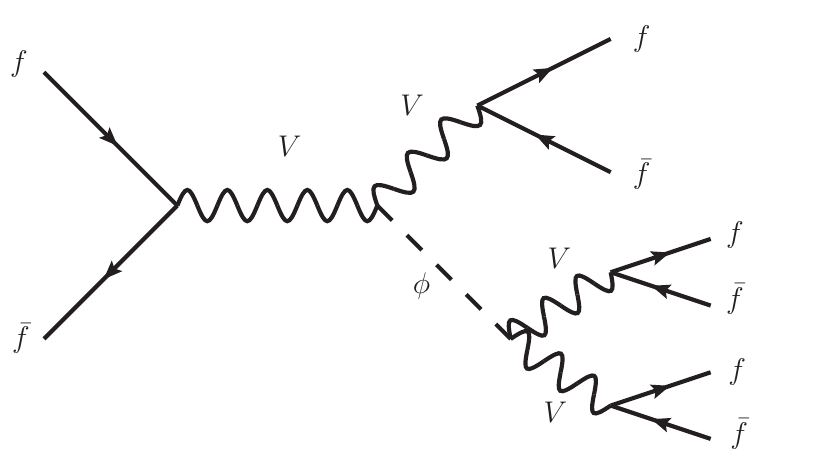}
         \caption{Dark Higgs-strahlung and decay into two dark photons.}
         \label{subfig:production_feynman:darkphoton_double}
     \end{subfigure}
     \begin{subfigure}[b]{0.32\textwidth}
         \centering
         \includegraphics[width=\textwidth]{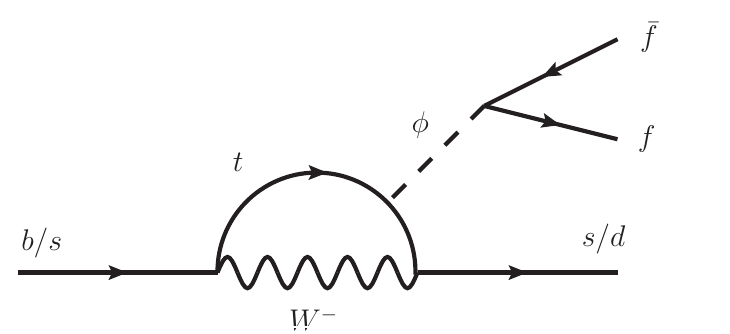}
         \caption{Rare $K$ and $B$ decays.\\ \quad}
         \label{subfig:production_feynman:mesondecay}
     \end{subfigure}

\caption{\label{fig:production_feynman} Feynman diagrams depicting the leading Dark Higgs search channels. Depending on the context, $f$ denotes either an SM fermion or a DM particle, and $V$ denotes either an SM gauge boson or a dark photon.}
\end{figure}

\subsection{Dark Higgs production}

Dark Higgs bosons that mix with the SM Higgs boson can be produced in proton-proton collisions in complete analogy to the SM Higgs boson. For heavy dark Higgs bosons, the dominant production mode is gluon fusion (see figure~\ref{subfig:production_feynman:gg_4f}), but other modes, such as vector boson fusion or production in association with gauge bosons or heavy quarks (see figure~\ref{subfig:production_feynman:gg_darkscalar_toptop}), may be of interest depending on the final state under consideration. 
If the dark Higgs boson has a mass $m_\phi < m_h/2$ it can also be produced in decays of the SM-like Higgs boson (see figure~\ref{subfig:production_feynman:higgs_dark_scalar_dark_scalar}). The corresponding decay width is given by
\begin{equation}
\Gamma_{h\to\phi\phi} = \frac{(m_h^2 + 2 \, m_\phi^2)^2 \, \sin^2 2 \theta}{128 \pi \, m_h}  \left(1 - \frac{4 \, m_\phi^2}{m_h^2}\right)^{1/2} \left(\frac{1}{w} \cos \theta + \frac{1}{v} \sin \theta \right)^2 \; .
\label{eq:Gammaphiphi}
\end{equation}

We note that for $\theta \ll 1$ this decay width is proportional to $\theta^2 / w^2$, which for $m_\phi \ll m_h$ is proportional to $\lambda_{h\phi}^2$. This is a direct consequence of the fact that this decay mode is allowed even in the limit $w \to 0$, i.e.\ for unbroken $U(1)'$ gauge symmetry. To compare constraints on this decay mode to other constraints (which depend exclusively on $\sin \theta$) it is therefore necessary to assume a specific value of $w$. In the following, we will consider the benchmark choice $w = 100\,\mathrm{GeV}$, keeping in mind that smaller (larger) values of $w$ will enhance (suppress) constraints from $h \to \phi \phi$ for a fixed value of $\sin \theta$. We remind the reader that -- just as in the SM -- it is perfectly possible to have $m_\phi \ll w$, while the opposite case ($m_\phi \gg w$) generally leads to unitarity violation~\cite{Kahlhoefer:2015bea}.

\begin{figure}%
\centering
     \begin{subfigure}[b]{0.35\textwidth}
         \centering
         \includegraphics[width=\textwidth]{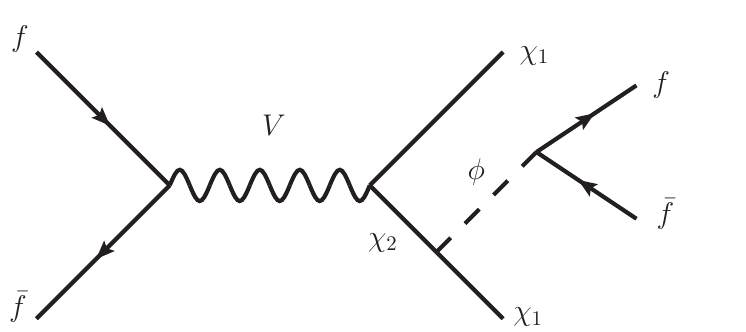}
         \caption{Inelastic dark matter.\\ \quad}
         \label{subfig:production_feynman:idm}
     \end{subfigure}
     \begin{subfigure}[b]{0.4\textwidth}
         \centering
         \includegraphics[width=\textwidth]{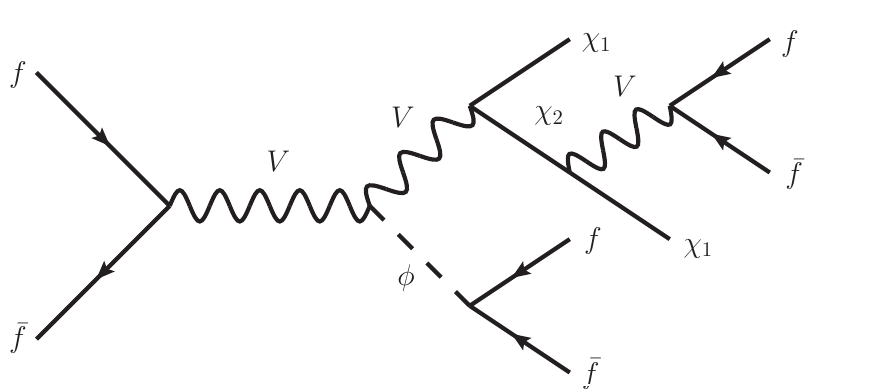}
         \caption{Dark Higgs-strahlung and inelastic dark matter.}
         \label{subfig:production_feynman:idmdh}
     \end{subfigure}
   
\caption{\label{fig:production_feynman:dm} Feynman diagrams depicting the leading Dark Higgs production channels in association with DM particles $\chi$.}
\end{figure}

If the dark Higgs field gives mass to an $A'$ gauge boson that couples to SM particles (either via direct charges or through kinetic mixing), it can furthermore be produced in association with the $A'$ in a process analogous to Higgs-strahlung:
\begin{equation}
 \text{SM} + \text{SM} \to A'^\ast \to A' + \phi \; ,
\end{equation}
see  figures~\ref{subfig:production_feynman:darkphoton} and~\ref{subfig:production_feynman:darkphoton_double}
This process is of particular interest if the $A'$ subsequently decays invisibly (e.g. into a pair of DM particles), such that the dark Higgs boson is produced in association with missing energy. 
The same signature may also be obtained from any process that produces DM particles through final-state radiation of a dark Higgs boson, provided that the DM particles obtain (part of) their mass from the dark Higgs field as depicted in Fig.~\ref{subfig:production_feynman:idm}.

Variations on the idea of dark Higgs-strahlung include the production of dark Higgs bosons in $Z$ boson decays ($Z \to A' + \phi$), which exploits the mass mixing between the dark photon and the $Z$ boson~\cite{Babu:1997st}, and the production of dark Higgs bosons in $\rho$ meson decays~\cite{Darme:2017glc}, which exploits the $\rho_0$--$\gamma$ mixing.

Finally, the strong experimental program searching for rare decays of kaons and $B$ mesons offers a unique opportunity to search for low-mass dark Higgs bosons. 
Indeed, dark Higgs bosons can participate in flavor-changing decays via the penguin diagram shown in figure~\ref{subfig:production_feynman:mesondecay}. 
The corresponding effective Lagrangian after electroweak symmetry breaking (and for $m_\phi < v$) can be written as~\cite{Batell:2009jf}
\begin{equation}
    \mathcal{L}_{bs} = h_{bs} \phi \bar{s}_L b_R + \text{h.c.} \; , \label{eq:LFCNC}
\end{equation}
where
\begin{equation}
    h_{bs} = \frac{\sin \theta \, m_b}{v} \frac{3 m_t^2 V_{ts}^\ast V_{tb}}{16\pi^2 v^2} \label{eq:hbs} 
\end{equation}
with $m_{b,t}$ being the quark masses and $V_{ij}$ denoting the CKM matrix elements.\footnote{Ref.~\cite{Kachanovich:2020yhi} identified a second contribution to $h_{bs}$ arising from the dark Higgs boson coupling to the longitudinal (i.e.\ Goldstone) mode of the $W$ boson in the loop. This contribution however turns out to be numerically sub-dominant.}
Given this effective coupling, a simple estimate of the inclusive decay rate $B \to X_s \phi$ can be obtained via~\cite{Winkler:2018qyg}\footnote{Early studies of the production of SM-like Higgs bosons in rare $B$ meson decays provide a similar expression with $m_B$ replaced by the quark mass $m_b$~\cite{Willey:1982mc,Chivukula:1988gp}. Although this expression predicts the wrong kinematic limit for dark Higgs boson production, it is still widely used in the community.}
\begin{equation}
    \Gamma_{B \to X_s \phi} = |h_{bs}|^2 \frac{(m_B^2 - m_\phi^2)^2}{32 \pi m_B^3} \; .
\end{equation}
Close to the kinematic threshold, i.e.\ $m_\phi \approx m_B - m_K$ a better estimate is obtained by considering separately the exclusive decays $B \to K \phi$ and $B \to K^\ast \phi$ with partial decay widths given by~\cite{Bobeth:2001sq}
\begin{align}
\Gamma(B \rightarrow K \, \phi) = & \frac{|h_{sb}|^2 }{64 \pi \, m_{B}^3} \lambda^{1/2}(m_{B}^2, m_{K}^2, m_\phi^2) \left| f_0^{B^0}(m_\phi^2)\right|^2 \left(\frac{m_{B}^2 - m_{K}^2}{m_b - m_s}\right)^2 \; , \\
\Gamma(B \rightarrow K^\ast \, \phi) = & \frac{|h_{sb}|^2  }{64 \pi \, m_{B}^3} \lambda^{3/2}(m_{B}^2, m_{K^\ast}^2, m_\phi^2) \left| A_0^{B^0}(m_\phi^2)\right|^2 \frac{1}{\left(m_b + m_s\right)^2}\; ,
\label{eq:Bwidth}
\end{align}
with $\lambda(a,b,c) =(a-b-c)^2-4\,b\,c$. The hadronic form factors are often parametrized as~\cite{Ali:1999mm,Ball:2004ye,Ball:2004rg}
\begin{align}
f_0^{B^0}(q^2) & = \frac{0.33}{1-\tfrac{q^2}{38\:\text{GeV}^2}} \; , \\
A_0^{B^0}(q^2) & = \frac{1.36}{1-\tfrac{q^2}{28\:\text{GeV}^2}} - \frac{0.99}{1-\tfrac{q^2}{37\:\text{GeV}^2}} \; .
\end{align}
More accurate numerical results for the hadronic form factors have recently been provided in Ref.~\cite{Gubernari:2018wyi}.
Even more exclusive final states have been considered in Ref.~\cite{Boiarska:2019jym}.

The effective Lagrangian for the transition $s \to d \phi$ can be obtained by replacing $b \to s$ and $s \to d$ in eqs.~\eqref{eq:LFCNC} and~\eqref{eq:hbs}. The resulting decay width of the kaon is given by~\cite{Leutwyler:1989xj,Bezrukov:2009yw}
\begin{align}
    \Gamma_{K^\pm \to \pi^\pm \phi} & = \frac{\lambda(m_K^2, m_\pi^2, m_\phi^2)^{1/2}}{m_K^3} \frac{|\mathcal{M}|^2}{16\pi} \\
    \Gamma_{K_L \to \pi^0 \phi} & = \frac{\lambda(m_K^2, m_\pi^2, m_\phi^2)^{1/2}}{m_K^3} \frac{\text{Re}(\mathcal{M})^2}{16\pi} \\
    \Gamma_{K^\pm \to \pi^\pm \phi} & = \frac{\lambda(m_K^2, m_\pi^2, m_\phi^2)^{1/2}}{m_K^3} \frac{\text{Im}(\mathcal{M})^2}{16\pi}
\end{align}
with
\begin{equation}
    \mathcal{M} \approx \frac{h_{sd}}{2} \frac{m_K^2 - m_\pi^2}{m_s - m_d} + \frac{7 \, \gamma_1 \sin \theta}{18}\frac{m_K^2 - m_\phi^2 + m_\pi^2}{v}
\end{equation}
and $\gamma_1 = 3.1 \times 10^{-7}$. The second term, arising from the effective Higgs-meson coupling, is numerically subleading and often neglected.

In principle, dark Higgs bosons can also be produced in $D$ meson decays. The corresponding effective coupling is however suppressed by small CKM matrix elements and the ratio $m_b^2 / m_t^2$, rendering it generally irrelevant for phenomenology. 

Finally, dark Higgs bosons may also be produced in proton bremsstrahlung through their effective coupling to nucleons (see section~\ref{sec:astro_and_dd}). This production mode may be of particular interest for forward experiments at the LHC~\cite{Foroughi-Abari:2021zbm}.

\subsection{Dark Higgs decays}

\begin{figure}%
\centering
     \begin{subfigure}[b]{0.6\textwidth}
         \centering
         \includegraphics[width=\textwidth]{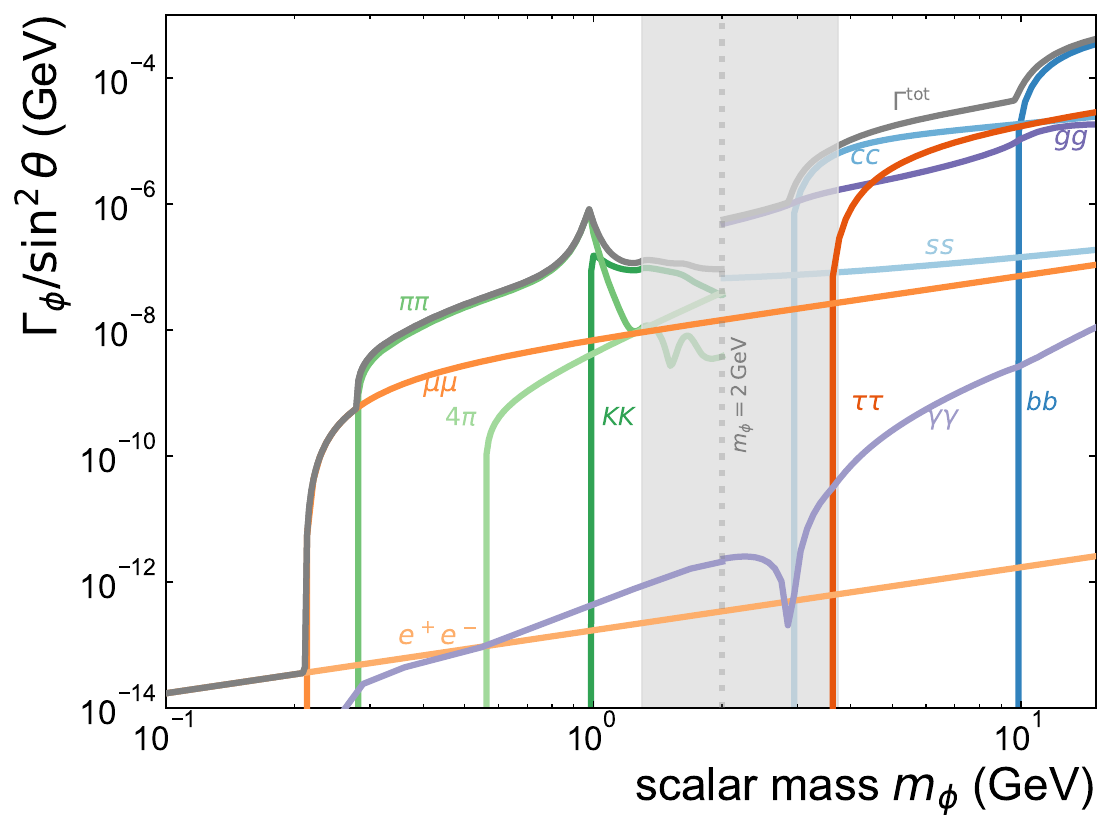}
         \caption{Light dark Higgs bosons.}
         \label{fig:gammaphi:light}
     \end{subfigure}
     \begin{subfigure}[b]{0.6\textwidth}
         \centering
         \includegraphics[width=\textwidth]{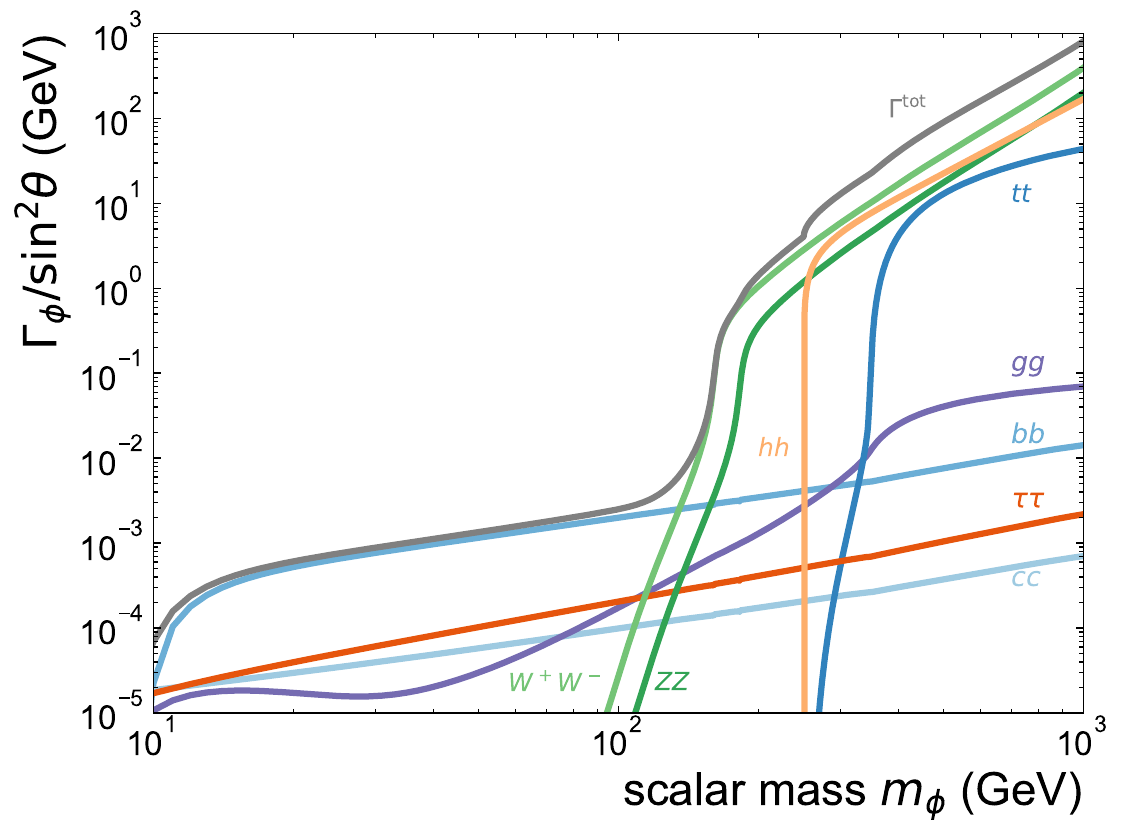}
         \caption{Heavy dark Higgs bosons.}
         \label{fig:gammaphi:heavy}
     \end{subfigure}
\caption{\label{fig:gammaphi} Decay widths for the dominant decay modes of (a) light and (b) heavy dark Higgs bosons. The gray shaded region in the top panel indicates the mass range where neither the dispersive analysis (following Ref.~\cite{Winkler:2018qyg}) nor the perturbative spectator model gives reliable predictions, leading to a discontinuity at $2 \, \mathrm{GeV}$. We emphasize that the hadronic partial decay widths in the GeV mass range are affected by large theoretical uncertainties and various approaches are used in the community. 
The decay widths in the bottom panel are taken from HDECAY~\cite{Djouadi:2018xqq}, except for $\Gamma_{\phi \to hh}$, for which we use the leading-order result from eq.~\eqref{eq:Gammahh} and approximate $\cos \theta \approx 1$.}
\end{figure}

The decay modes of a real scalar with Higgs-like couplings to SM particles have been studied extensively in the literature and are summarized in figure~\ref{fig:gammaphi}. For example, the leading-order partial decay widths into leptons are given by
\begin{equation}
    \Gamma_{\phi \to \ell^+ \ell^-} = \frac{\sin^2 \theta \, m_\phi \, m_\ell^2}{8\pi \, v^2} \sqrt{1 - \frac{4m_\ell^2}{m_\phi^2}} \; .
\end{equation}
The public tool HDECAY~\cite{Djouadi:2018xqq} provides partial decay widths for SM-like Higgs bosons with a mass in the range $2 m_D \lesssim m_\phi \lesssim 1 \, \mathrm{TeV}$ including many relevant higher-order corrections.\footnote{We emphasize that HDECAY does not account for the reduced phase space due to confinement, i.e.\ the decay into charm (bottom) quarks opens up for $m_\phi > 2 m_c$ ($m_\phi > 2 m_b$) rather than $m_\phi > 2 m_D$ ($m_\phi > 2 m_B$). To first approximation, this effect can be included by multiplying the decay widths from HDECAY with a correction factor of the form $\sqrt{1 - 4 m_{D,B}^2 / m_\phi^2} / \sqrt{1 - 4 m_{c,b}^2 / m_\phi^2}$.} For even larger masses, the decay width becomes unphysical due to diverging next-to-leading order electroweak corrections, and one should revert to the tree-level decay widths instead~\cite{Bringmann:2021sth}. 

For masses below the D meson threshold, it becomes essential to take into account the confinement of the final-state particles. For $m_\phi \lesssim 1.3 \, \mathrm{GeV}$ relatively accurate estimates for the decays into $\pi\pi$ and $KK$ can be obtained using dispersion relations~\cite{Donoghue:1990xh}, while the decay into photons can be deduced from low-energy theorems~\cite{Leutwyler:1989tn}. For higher masses, additional final states become important and an accurate prediction of the various decay widths becomes very challenging. So far no fully satisfactory agreement has been achieved in the literature. The most widely adopted approach is the one from Ref.~\cite{Winkler:2018qyg}, which switches from a dispersive analysis to a perturbative spectator model at 2 GeV (see the top panel of figure~\ref{fig:gammaphi}). We note, however, that this approach has been questioned in Refs.~\cite{Bezrukov:2018yvd,Gorbunov:2023lga}, where it has been argued that for $m_\phi$ close to the $f_0(980)$ resonance the partial decay width into $2\pi$ is overestimated (see also Ref.~\cite{Monin:2018lee}).

A notable deviation from the expectations for an SM-like Higgs boson arises for $m_\phi > 2 m_h$ (see the bottom panel of figure~\ref{fig:gammaphi}). In this case, the dark Higgs boson can decay into a pair of SM Higgs bosons with a decay width given by
\begin{equation}
\Gamma_{\phi \to h h} = \frac{(m_\phi^2 + 2 \, m_h^2)^2 \, \sin^2 2 \theta}{128 \pi \, m_\phi}  \left(1 - \frac{4 \, m_h^2}{m_\phi^2}\right)^{1/2} \left(\frac{1}{v} \cos \theta + \frac{1}{w} \sin \theta \right)^2 \; .
\label{eq:Gammahh}
\end{equation}
This decay mode is of particular interest in the context of searches for Higgs boson pair production at the LHC.

In the presence of a dark photon, there are two additional decay modes of interest for phenomenology. The first is the tree-level decay $\phi \to A' A'$ followed by the decay of each dark photon into SM fermions. 
If $m_\phi > 2 m_{A'}$ the dark photons are on-shell and the decay width is given by~\cite{Batell:2009yf}
\begin{equation}
 \Gamma_{\phi \to A' A'} = \frac{g'^2 m_\phi^3}{32\pi m_{A'}^2}  \sqrt{1 - \frac{4 m_{A'}^2}{m_\phi^2}} \left(1 - \frac{4 m_{A'}^2}{m_\phi^2} + \frac{12 m_{A'}^4}{m_\phi^4} \right) \; , 
\end{equation}
where $g'$ denotes the $U(1)'$ gauge coupling.
If all four SM fermions are detected, it is possible to reconstruct both the dark photon and the dark Higgs mass from the final state. 
If the decay into on-shell dark photons is kinematically forbidden, it becomes interesting instead to consider the loop-induced decay into two SM fermions~\cite{Batell:2009yf}:
\begin{equation}
 \Gamma_{\phi\to f \overline{f}} = \frac{3 \, g_f^4 \, g'^2 \, m_\phi}{128\pi^5} \frac{m_f^2}{m_{A'}^2} \left(1-\frac{4 m_f^2}{m_\phi^2}\right)^{3/2} \left| I\left(\frac{m_\phi^2}{m_{A'}^2},\frac{m_f^2}{m_{A'}^2}\right) \right|^2 \; ,
\end{equation}
where $g_f$ denotes the coupling of the dark photon to SM fermions and
\begin{equation}
 I(x_\phi, x_f) \equiv \int_0^1 \mathrm{d}y \int_0^{1-y} \mathrm{d}z \frac{2-(y+z)}{(y+z)+(1-y-z)^2 x_f - y z x_\phi} \approx \frac{3}{2}
\end{equation}
for $m_{A'} \gg m_\phi, m_f$. This decay mode will typically only be relevant if the mixing with the SM Higgs boson is extremely small~\cite{Darme:2017glc}, but may give rise to interesting phenomenology, such as long-lived dark Higgs bosons~\cite{Bernreuther:2020xus}. We also note that an interesting feature of the loop-induced decay into SM fermions is that the branching ratios may differ from the ones obtained through Higgs mixing. For example, for a dark photon with kinetic mixing, the partial decay widths into the various quarks are proportional to $q_q^4 m_q^2$, where $q_q$ denotes the electromagnetic charge of quark $q$. In most of the parameter space, decays into charm quarks would therefore dominate over decays into bottom quarks.

\subsection{Dark Higgs decay length}

The distance $d$ traveled by a particle with lifetime $\tau_\phi$, mass $m_\phi$ and momentum $p_\phi$ before decaying follows an exponential distribution
\begin{equation}
    P(d) = \frac{1}{l} \exp(- d/l)
\end{equation}
with $l = c \tau_\phi p_\phi / m_\phi$. As long as $l$ is small compared to the typical vertex resolution of a given experiment, the majority of decays will appear prompt. Conversely, if $l$ is large compared to the size of the detector, most particles will decay outside the detector and hence appear invisible. In the intermediate regime, it may be possible to reconstruct a displaced decay vertex, offering a powerful handle for background suppression.

For concreteness, let us consider two examples. A dark Higgs boson with $m_\phi = 200\,\mathrm{MeV}$ may be produced in $B$ meson decays with $p_\phi \approx m_B / 2$. In this scenario, we find
\begin{equation}
    l \approx \frac{7 \, \mathrm{cm}}{\sin^2 \theta} \; ,
\end{equation}
such that for $\sin \theta \lesssim 0.1$ the dark Higgs boson is expected to escape from the detector unnoticed in a large fraction of events.

As a second example, let us consider a dark Higgs boson with $m_\phi = 500 \, \mathrm{MeV}$ produced in a 400~\gev proton beam-dump experiment. Given the typical momentum $p_\phi \approx 10 \, \mathrm{GeV}$~\cite{Bezrukov:2009yw}, we find
\begin{equation}
    l \approx 20 \, \mathrm{cm} \left(\frac{10^{-3}}{\sin \theta}\right)^2 \; .
\end{equation}

\FloatBarrier
\section{Constraints from observations of the SM-like Higgs boson}
\label{sec:Signal_strength}
\subsection{Signal strength}

Both ATLAS and CMS search extensively for non-standard couplings of the SM-like Higgs boson. The simplest extension is to consider a common signal strength modifier $\mu$ affecting equally all production and decay modes of the SM-like Higgs boson, keeping all branching fractions equal to the SM predictions. In the case of mixing with a dark Higgs boson, the signal strength of the SM-like Higgs boson is suppressed according to~\cite{Bhattiprolu:2022ycf}
\begin{equation}
\mu = \cos^2 \theta \frac{\Gamma^\text{SM}_h \cos^2 \theta}{\Gamma^\text{SM}_h \cos^2 \theta + \Gamma_{h \to \text{dark}}} \; , 
\end{equation}
where $\Gamma^\text{SM}_h$ denotes the decay width of the SM Higgs boson and $\Gamma_{h \to \text{dark}}$ denotes the decay width for any decays of the SM-like Higgs boson into dark sector states, irrespective of the experimental signature. If all dark sector states are heavier than $m_h/2$, this expression reduces to $\mu = \cos^2 \theta$. The latest bounds on $\mu$ are given by~\cite{CMS:2022dwd,ATLAS:2022vkf}
\begin{equation}
    \mu = \begin{cases}
1.05 \pm 0.06 & \; \text{(ATLAS)} \\
1.00 \pm 0.06 & \; \text{(CMS)}.
\end{cases}
\end{equation}
Rather than attempting to combine these two measurements considering all correlations, we will use the average of the two results and assume that the dominant uncertainties are of systematic nature. The resulting signal strength of $1.03 \pm 0.06$ corresponds to 
\begin{equation}
    \sin \theta < 0.27 
\end{equation}
at 95\% confidence level (CL) in the case of no dark decays. 

\subsection{Invisible decays}

Additional constraints arise in the case that the SM-like Higgs boson can decay into fully invisible final states. In our set-up, such decays arise if the SM-like Higgs boson can decay into a pair of dark Higgs bosons ($h \to \phi \phi$) and both dark Higgs bosons decay invisibly or escape from the detector before decaying, or if the SM-like Higgs boson inherits the couplings of the dark Higgs boson to other dark sector states. Experimental constraints are quoted in terms of
\begin{equation}
    \mathcal{B}_\text{inv} = \frac{\sigma}{\sigma_\text{SM}}\frac{\Gamma_{h \to \text{inv}}}{\Gamma_{h}} \; ,
\end{equation}
which in our case corresponds to
\begin{equation}
    \mathcal{B}_\text{inv} = \cos^2 \theta \frac{\Gamma_{h \to \text{inv}}}{\Gamma^\text{SM}_h \cos^2 \theta + \Gamma_{h \to \text{dark}}} \; .
\end{equation}
The leading constraints on $\mathcal{B}_\text{inv}$ stem from Higgs production in vector boson fusion and yield~\cite{ATLAS:2022yvh,CMS:2022qva}
\begin{equation}
\mathcal{B}_\text{inv} < \begin{cases}
0.15 & \; \text{(ATLAS)} \\
0.18 & \; \text{(CMS)}
\end{cases} \; 
\end{equation}
at 95\% confidence level. An even stronger bound is obtained by combining different production modes, giving $\mathcal{B}_\text{inv} < 0.11$~\cite{ATLAS:2020kdi}. For $m_\phi \ll m_h /2$ and assuming invisible decays of the dark Higgs boson, we find
\begin{equation}
    \mathcal{B}_\text{inv} \approx 0.05 \left(\frac{\sin \theta}{0.01}\right)^2
\end{equation}
for $w = 100 \, \mathrm{GeV}$, and hence $\sin \theta \lesssim 0.02$. On the other hand, using $\mu = \cos^2 \theta - \mathcal{B}_\text{inv}$, the measurement of the Higgs signal strength discussed above gives $\sin \theta \lesssim 0.01$. As pointed out in Ref.~\cite{Biekotter:2022ckj}, this bound is always stronger than the ones obtained from invisible Higgs decays, so we will not consider the latter further in the following.

\subsection{Future projections and proposed experiments}

The relative importance of the Higgs signal strength measurement and  searches for invisible Higgs decays is expected to change with HL-LHC. Indeed, Ref.~\cite{Cepeda:2019klc} finds that the bound on $\mu$ will only improve significantly if systematic uncertainties can be reduced, and even under optimistic assumptions will only reach an expected 95\% CL lower bound of $\mu > 0.96$, assuming that the best-fit value agrees with the SM. The upper bound on the invisible branching ratio, on the other hand, will continue to improve substantially, with an expected upper bound of $\mathcal{B}_\text{inv} < 0.025$ at 95\% CL with an integrated luminosity of $3000\,\mathrm{fb^{-1}}$.

Future electron-positron colliders will provide tremendous progress in constraining the properties of the SM-like Higgs boson. In particular, they will provide very accurate measurements of the $Zh$ coupling, which should constrain $\mu$ at the level of $0.005$~\cite{deBlas:2019rxi}, corresponding to an expected upper bound $\sin \theta \lesssim 4 \times 10^{-3}$ at 95\% CL~\cite{EuropeanStrategyforParticlePhysicsPreparatoryGroup:2019qin}, while the expected upper bound on the invisible branching ratio is $\mathcal{B}_\text{inv} \lesssim 0.002$. As for HL-LHC, the latter will give the dominant bound on light dark Higgs bosons, giving $\sin \theta \lesssim 2 \times 10^{-3}$~\cite{Liu:2017lpo}.

\FloatBarrier
\section{Direct dark Higgs searches}
\label{sec:direct_Higgs}
In this section, we consider searches for dark Higgs bosons that do not rely on any other state in the dark sector, i.e.\ the dark Higgs boson is produced directly from SM states and decays back into SM states. These searches are hence reminiscent of the long search for an SM-like Higgs boson at colliders, except that for the case of the dark Higgs boson the coupling strength is unknown. Moreover, the SM-like Higgs boson can now appear as the initial or final state, participating in dark Higgs production or decay.

\begin{figure}[t]%
\centering
\includegraphics[height=210px]{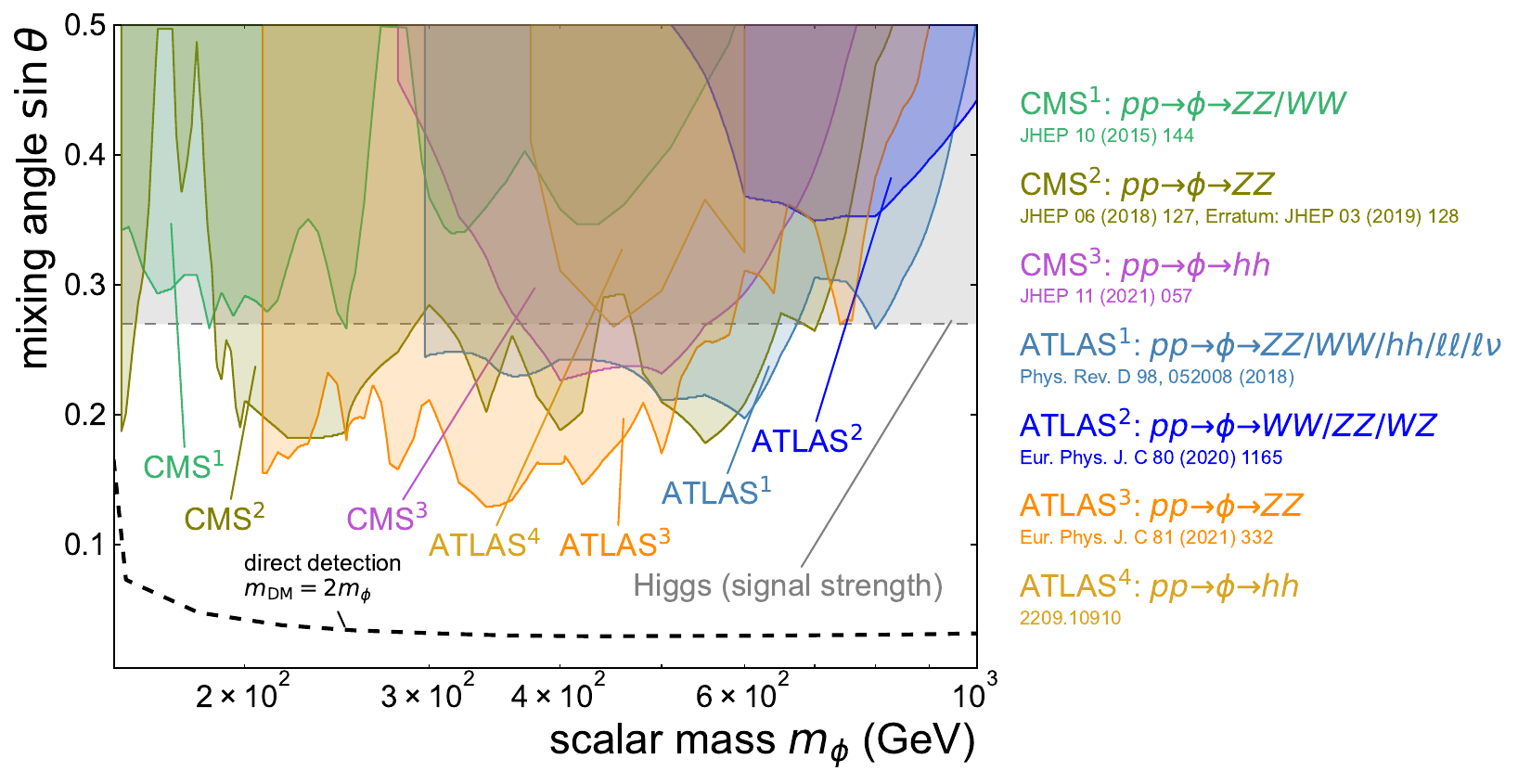}
\caption{\label{fig:direct_high} \hbox{95\,\% CL} upper limits on the mixing angle $\sin\theta$ as function of scalar mass $m_S$ for heavy dark Higgs bosons from CMS~\cite{CMS:2015hra,CMS:2018amk,CMS:2021yci} and ATLAS~\cite{ATLAS:2018sbw,ATLAS:2020fry,ATLAS:2020tlo,ATLAS:2022xzm}. 
For limits from Higgs signal strength see Sec.~\ref{sec:Signal_strength}.
For limits from direct detection experiments see Sec.~\ref{sec:astro_and_dd}.
Constraints colored in gray with a dashed outline are reinterpretations not performed by the experimental collaborations and without access to raw data.
}
\end{figure}

\begin{figure}[t]%
\centering
\includegraphics[height=210px]{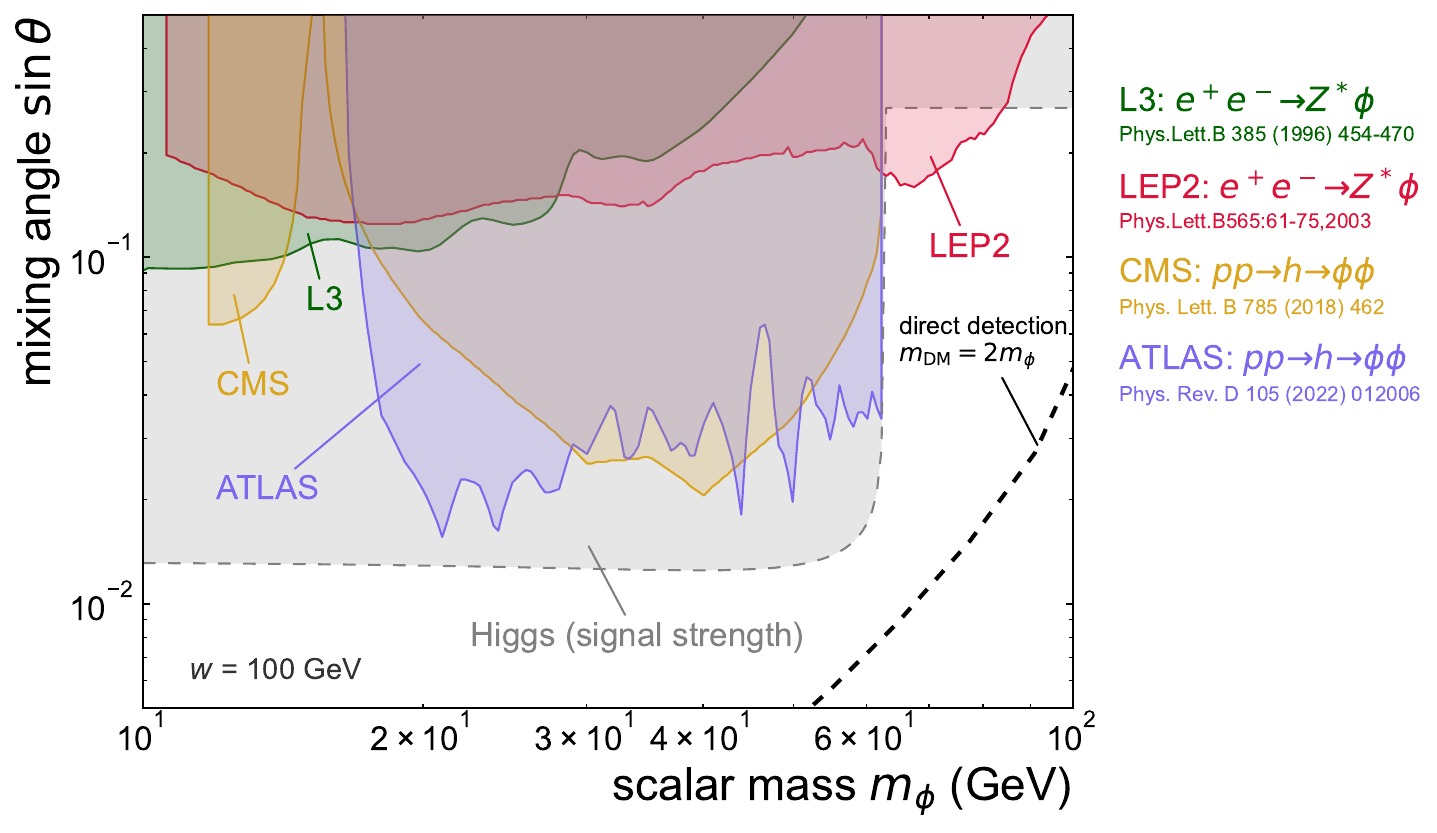}
\caption{\label{fig:direct_medium} \hbox{95\,\% CL} upper limits on the mixing angle $\sin\theta$ as function of scalar mass $m_S$ for light dark Higgs bosons from CMS~\cite{CMS:2018zvv}, ATLAS~\cite{ATLAS:2021hbr}, LEP2~\cite{LEPWorkingGroupforHiggsbosonsearches:2003ing}, and L3~\cite{L3:1996ome}. 
For limits from Higgs signal strength see Sec.~\ref{sec:Signal_strength}.
For limits from direct detection experiments see Sec.~\ref{sec:astro_and_dd}.
Constraints colored in gray with a dashed outline are reinterpretations not performed by the experimental collaborations and without access to raw data.}
\end{figure}

\begin{figure}[t]
\centering
\includegraphics[height=210px]{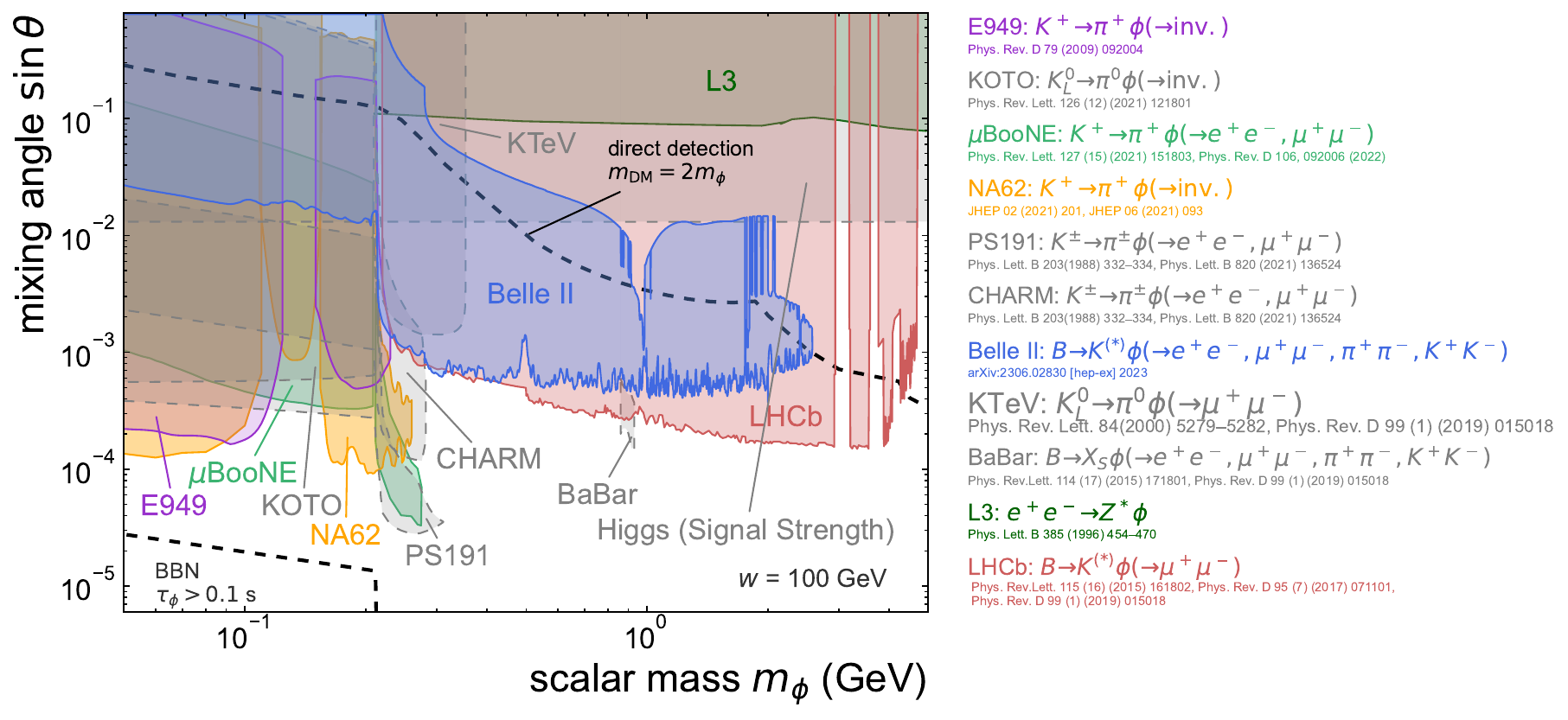}%
\caption{\label{fig:direct_light}  Upper  limits on the mixing angle $\sin\theta$ as function of scalar mass $m_S$ from
PS191\,(\hbox{90\,\% CL}\,\cite{Gorbunov:2021ccu}),
E949\,(\hbox{90\,\% CL}\,\cite{BNL-E949:2009dza}),
NA62\,(\hbox{90\,\% CL}\,\cite{NA62:2020pwi,NA62:2021zjw}),
MicroBooNE\,(\hbox{95}\,\% CL\,\cite{MicroBooNE:2021usw,MicroBooNE:2022ctm} and this work),
KOTO\,(\hbox{90\,\% CL}\,\cite{KOTO:2020prk} and this work), 
KTeV\,(\hbox{90\,\% CL}\,\cite{KTEV:2000ngj}),
L3\,(\hbox{95\,\% CL}\,\cite{L3:1996ome} and this work),
CHARM\,(\hbox{95\,\% CL}\,\cite{CHARM:1985anb, Winkler:2018qyg}),
\lhcb\,(\hbox{95\,\% CL}\,\cite{LHCb:2015nkv, LHCb:2016awg, Winkler:2018qyg}),
\belletwo\,(\hbox{95\,\% CL}\,\cite{Belle-II:2023ueh}),
\babar\,(\hbox{90\,\% CL}\,\cite{BaBar:2015jvu,Winkler:2018qyg}).
For limits from Higgs signal strength see Sec.~\ref{sec:Signal_strength}.
For limits from BBN see Sec.~\ref{sec:cosmo_and_indirect}.
For limits from direct detection experiments see Sec.~\ref{sec:astro_and_dd}.
Constraints colored in gray with a dashed outline are reinterpretations not performed by the experimental collaborations and without access to raw data.}
\end{figure}

We note that in the absence of other dark sector states the dark Higgs boson can be mapped onto the more general class of dark scalar models first introduced in Ref.~\cite{OConnell:2006rsp}. In these models the linear coupling $\mu \scalar H^\dagger H$ and the quadratic coupling $\lambda \scalar^2 H^\dagger H$ are allowed to vary independently, whereas for the dark Higgs boson, they are related via $\mu = 2 w \lambda$ due to the underlying symmetry. Nevertheless, in both settings, the leading effect is parametrized by the effective Higgs mixing angle $\sin \theta$, for which constraints can be obtained as a function of $m_\scalar$. We show a summary of these constraints in figures~\ref{fig:direct_high},~\ref{fig:direct_medium} and~\ref{fig:direct_light} and discuss the individual searches in detail below. Unless explicitly stated otherwise, the constraints discussed in sections~\ref{subsec:limits_collider_heavy} and \ref{subsec:limits_collider_medium} have been translated to the parameter space of the dark Higgs model using HiggsBounds~\cite{Bechtle:2020pkv} as implemented in HiggsTools~\cite{Bahl:2022igd} and are shown at 95\% confidence level (CL). We focus on searches and results that give the strongest constraints on dark Higgs bosons. We do not consider the case where the dark Higgs boson is close in mass to the SM-like Higgs boson and therefore exclude the mass range $100 \gev \leq m_\phi \leq 150 \gev$, which has recently been studied in great detail in Ref.~\cite{Bhattiprolu:2022ycf}.

\subsection{Collider limits for \texorpdfstring{$\mscalar > m_h$}{ms greater than mh}}
\label{subsec:limits_collider_heavy}

\subsubsection*{\texorpdfstring{$ pp \to \scalar \to ZZ/WW$}{pp to phi to ZZ or WW} at CMS}
\label{subsec:limits_collider_heavy_1}
In 2011 and 2012 the LHC collided protons at a center-of-mass energy of $7$ resp. $8 \tev$. The CMS collaboration used these data corresponding to an integrated luminosity of up to $5.1 \fbinv$ and up to $19.7 \fbinv$ to search for heavy Higgs bosons decaying via a pair of $W$ or $Z$ bosons into final states with one, two, or four charged leptons~\cite{CMS:2015hra}. In the case of four charged leptons, one Z boson is allowed to decay into a pair of tau leptons, while otherwise leptons are restricted to electrons and muons. 
The analysis excludes Higgs bosons with SM-like couplings in the mass range between 145 and $1000 \gev$ at the 95\% CL. These results can be translated into limits on the mixing angle. With an upper bound of $\sin \theta \lesssim 0.26$, the strongest limits are reached in the mass range below approximately $260 \gev$.

\subsubsection*{\texorpdfstring{$pp \to \scalar\to ZZ$}{pp to phi to ZZ} at CMS}
\label{subsec:limits_collider_heavy_2}
Higgs masses up to $3 \tev$ are probed by the CMS collaboration in an analysis based on $pp$ collisions at $\sqrts = 13 \tev$ recorded in 2016, and corresponding to an integrated luminosity of $35.9 \fbinv$~\cite{CMS:2018amk}. The search considers both gluon fusion as well as electroweak production of a heavy Higgs boson $\scalar$ decaying into a pair of $Z$ bosons. Interference effects between a resonant signal of arbitrary width and background amplitudes are included. $ZZ$ decays are reconstructed using the 4$\ell$, 2$\ell$2q, and 2$\ell$2$\nu$ final states, where $q$ denotes a quark leading to a jet in the final state. 
In order to categorize events according to their production mechanism or to separate the signal from the dominant backgrounds, all relevant matrix element probabilities are calculated for each event and compared to each other. The analysis provides great sensitivity in the mass range between 200 and $600 \gev$ and the strongest upper bound of $\sin \theta \lesssim 0.17$ is reached for resonance masses around $550 \gev$.   

\subsubsection*{\texorpdfstring{$pp \to \scalar\to ZZ/WW/hh/\ell\ell/\ell\nu$}{pp to phi to ZZ or WW or hh or ellellellnu} at ATLAS}
\label{subsec:limits_collider_heavy_3}
A search for a heavy Higgs boson decaying into a pair of $Z$, $W$, or Higgs bosons as well as directly into leptons has been published by ATLAS~\cite{ATLAS:2018sbw}. The analysis makes use of $pp$ collisions at $\sqrts = 13 \tev$ corresponding to a total integrated luminosity of $36.1 \fbinv$. 
To separate the SM background from a potential signal, the invariant final state mass distribution is examined. While the signal shape is extracted from Monte Carlo~(MC) simulations, the background shape and normalization are estimated using a mixture of MC simulation and data from dedicated control regions. The final result is based on the combination of 12 different final states from the bosonic decay channels as well as the $\ell \nu$ and $\ell \ell$ final states. 
Above about $600 \gev$, this search provides the strongest direct limits over a wide mass range with the best upper bound of $\sin \theta \lesssim 0.2$ at $600 \gev$. 

\subsubsection*{\texorpdfstring{$pp \to \scalar \to WW/ZZ/WZ$}{pp to phi to WW or ZZ or WZ} at ATLAS}
\label{subsec:limits_collider_heavy_4}
The first search for heavy Higgs bosons based on the full LHC $pp$ run at $13 \tev$ has been published by ATLAS~\cite{ATLAS:2020fry}. The data corresponding to a total integrated luminosity of $139 \fbinv$ were recorded between 2015 and 2018. The search for $\scalar \to WW/ZZ/WZ$ is performed in final states in which one boson decays leptonically, and the other hadronically. 
To discriminate signal from the background, the reconstructed (transverse) mass of the $VV$ system is used when the leptonically decaying $V$ boson decays into a pair of leptons (neutrinos). 
As the analysis adapts the reconstruction of the hadronically decaying $V$ boson for high transverse momenta, the search is particularly sensitive above around $800 \gev$. However, with the best upper limit of $\sin \theta \lesssim 0.35$, this search is less sensitive than indirect limits from measurements of the Higgs signal strength.   

\subsubsection*{\texorpdfstring{$pp \to \scalar\to ZZ$}{pp to phi to ZZ} at ATLAS}
\label{subsec:limits_collider_heavy_5}
The strongest limits on the mixing angle are provided by the ATLAS experiment~\cite{ATLAS:2020tlo}. Using the complete LHC $13 \tev$ $pp$ dataset recorded between 2015 and 2018, ATLAS published a search for heavy Higgs bosons decaying via a pair of $Z$ bosons into $4 \ell$ and $2\ell 2\nu$ final states. The analysis looks for an excess in the invariant mass of the four charged leptons respectively the transverse mass of the two charged leptons and the two neutrinos. The results are interpreted separately for the gluon-gluon and vector-boson fusion production modes. At about $340 \gev$, the best constraints are reached with an upper bound of $\sin \theta \lesssim 0.13$.

\subsubsection*{\texorpdfstring{$pp \to \scalar\to hh $}{pp to phi to hh} at CMS}  
\label{subsec:limits_collider_heavy_6}
Using an integrated luminosity of $137 \fbinv$ of $13 \tev$ $pp$ collisions recorded between 2016 and 2018, CMS performed a search for heavy Higgs bosons from gluon-gluon fusion decaying into the SM Higgs boson and another Higgs boson~\cite{CMS:2021yci}. The mass of both BSM Higgs bosons is a free parameter of the search. 
To maximize the statistics while still having a clean signature, the analysis explores final states with two $\tau$ leptons and two $b$ quarks. A fully connected, feed-forward neural network is used to classify events into five signal and background-enriched categories per year of data-taking and per $\tau$ decay channel. The categories enriched in background events are used to constrain systematic uncertainties. 
For our re-interpretation, the heavy Higgs boson is associated with the dark Higgs boson, while the second Higgs boson from the decay is assumed to be the SM Higgs boson. 
With an upper bound of $\sin \theta \lesssim 0.23$, the best sensitivity of this search is reached between 400 and $500 \gev$.  

\subsubsection*{\texorpdfstring{$pp \to \scalar \to  h h$}{pp to phi to hh} at ATLAS} 
\label{subsec:limits_collider_heavy_7}
Similar to the $pp \to \scalar\to hh $ search performed at the CMS experiment, the ATLAS collaboration searched for Higgs boson pair production in final states with two $b$-quark jets and two $\tau$ leptons using an integrated luminosity of $139 \fbinv$~\cite{ATLAS:2022xzm}. The analysis targets off-shell and scalar, narrow-resonant production in the mass range between $250$ and $1600 \gev$. At least one of the $\tau$ leptons is required to decay hadronically. Multivariate discriminants are used to reject and constrain background events and to extract the $hh$ signal yields in a binned maximum likelihood fit.  
Since this search is not yet available in the public version of HiggsBounds, we perform the translation ourselves, using the production cross-section of an SM-like Higgs boson from Ref.~\cite{LHCHiggsCrossSectionWorkingGroup:2016ypw}. As in the case of the CMS result, the best sensitivity is reached between 400 and $500 \gev$ with a best upper bound of $\sin \theta \lesssim 0.26$. 

\FloatBarrier
\subsection{Collider limits for \texorpdfstring{$10~\gev \leq \mscalar\leq m_h$}{10 GeV less than ms less than mh}}
\label{subsec:limits_collider_medium}

\subsubsection*{\texorpdfstring{$e^+e^-\to Z^*\scalar$}{ee to Zphi} at \lthree}
\label{subsec:limits_collider_light_l3}
Using an integrated luminosity of $114 \pbinv$ the \lthree experiment at the $e^+e^-$ collider LEP has searched for $e^+e^-\to Z^*\scalar$, combining events with $Z^*$ decays into $\nu\bar{\nu}$, $e^+e^-$ and $\mu^+\mu^-$ and $\scalar$ reconstructed as one, two and three jets depending on the $\scalar$ mass~\cite{L3:1996ome}.
For $\mscalar > 2m_{\mu}$, the $\scalar$ decays promptly.
We reinterpret the 95\,\%~CL limits derived by \lthree using all $Z^*$ decay channels to constrain scalar masses $2m_{\mu} < \mscalar < 60\,\gev$ at the level of $\sin\theta = \sqrt{\Gamma(Z\to Z^*\scalar)/\Gamma^{SM}(Z\to Z^*H)} \lesssim 0.1$.
For scalar masses $\mscalar < 2m_{\mu}$ the decay length of the dark Higgs boson becomes macroscopic, such that the assumption of prompt $\scalar$ decays made in the L3 analysis is not satisfied for a large fraction of events.
We do not attempt a reinterpretation that would require assumptions on the \lthree detector response in this mass range, since several stronger limits from beam-dumps, fixed target experiments, and \belletwo exist.

\subsubsection*{\texorpdfstring{$e^+e^-\to Z^*\scalar$}{ee to Zphi} at LEP2}
\label{subsec:limits_collider_light_lep2}
Similarly to the \lthree result, the limits of the combined search for an SM Higgs boson of the four LEP experiments~\cite{LEPWorkingGroupforHiggsbosonsearches:2003ing} can be translated into limits on the dark Higgs boson. Using an integrated luminosity of up to $2461 \pbinv$ of $e^+e^-$ data at center-of-mass energies between 189 and $209 \gev$, the LEP experiments searched for $e^+e^-\to Z^*\scalar$ in final states with either four jets $(\scalar \to b \bar{b}) (Z \to q \bar{q})$, or missing energy $(\scalar \to b \bar{b})( Z \to \nu \bar{\nu})$, or two charged electrons or muons $(\scalar \to b \bar{b})( Z \to \ell^+ \ell^-)$, or two $\tau$ leptons $(\scalar \to b \bar{b}) (Z \to \tau^+\tau^-)$ and $(\scalar \to \tau^+ \tau^-) (Z\to q \bar{q})$. In most channels, the input is binned in two variables: The reconstructed Higgs boson mass as well as a variable that combines information from $b$-tagging with neural network outputs of high-level event features. 
Between 10 and $80 \gev$, the obtained exclusion limits for $\sin\theta$ range from 0.15 to 0.2.

\subsubsection*{\texorpdfstring{$pp \to h \to \scalar\scalar$}{pp to h to phiphi} at CMS}
\label{subsec:limits_collider_light_cms}

Limits on the production cross-section of the SM Higgs boson decaying into a pair of light pseudoscalar bosons can directly be translated into limits on the production of light dark Higgs bosons from SM Higgs decays. A first such search in final states with two $b$ quarks and two $\tau$ leptons has been performed by CMS in 2016 using a total integrated luminosity of $35.9 \fbinv$ of $13 \tev$ $pp$ collisions data~\cite{CMS:2018zvv}.  
 Four categories are defined based on the invariant mass of the leading $b$-tagged jet and the visible decay products of the $\tau$ leptons. For signal events, this variable is bound from above by the mass of the SM Higgs boson, while it is on average much larger for background events as the three objects do not originate from a resonance decay. The final result is obtained from a fit of the visible $\tau \tau$ mass distribution in each category. With upper limits of $\sin \theta \lesssim 0.02$, this search provides the strongest direct limits on low-mass dark Higgs bosons in a mass range between 30 and $45 \gev$. 

\subsubsection*{\texorpdfstring{$pp \to h \to \scalar\scalar$}{pp to h to phiphi} at ATLAS}
\label{subsec:limits_collider_light_atlas}
 Using all 13~\tev $pp$ collision data from the LHC recorded between 2015 and 2018, ATLAS performed a search for decays of the SM Higgs boson into a pair of new pseudoscalar particles in final states with two muons and two $b$-quarks~\cite{ATLAS:2021hbr}. A narrow dimuon resonance is searched for in the invariant mass spectrum between $16$ and $62 \gev$. Boosted decision tree techniques are used to separate the signal from the background. 
Compared to the $ pp \to h \to \scalar\scalar$ CMS search discussed above, this analysis achieves significantly higher sensitivities in the mass range between 20 and 30~\gev, with exclusions reaching down to $\sin \theta \lesssim 0.015$.

\subsection{Collider limits for \texorpdfstring{$50~\mev \leq \mscalar\leq 5~\gev$}{50 MeV less than ms less than 5 GeV}}
\label{subsec:limits_collider_low}

\subsubsection*{\texorpdfstring{$B \to K^{(*)}\scalar(\to e^+e^-, \mu^+\mu^-, \pi^+\pi^-, K^+K^-)$}{B to K phi to ee, mumu, pipi, or KK} at \belletwo}
\label{subsec:limits_collider_light_belletwo}

\belletwo at the asymmetric $e^+e^-$ collider SuperKEKB in Japan has searched directly for long-lived spin-0 mediators emerging from $b\to s$ transitions in \bmeson decays~\cite{Belle-II:2023ueh} utilizing an integrated luminosity of 189~\fbinv.
The search has been conducted model-independently in eight exclusive final states using $K$ and \kstarz-mesons and mediator decays into $e^+e^-$, $\mu^+\mu^-$, $\pi^+\pi^-$, and $K^+K^-$ for various lifetimes but excluding prompt decays.
These are so far the only exclusive limits for scalar decays into hadrons.
Using a combined fit, the results are also presented as 95\,\%~CL limits in the \mscalar--$\sin \theta$ plane and constrain the mixing angle for dark Higgs boson masses $25\,\mev < m_{\scalar} \lesssim 2.5\,\gev$.
For $m_{\scalar} > 2.0\,\gev$ only decays into muons are used in the combined fit to avoid large uncertainties in the branching fraction calculations into two hadrons.
Note that in contrast to searches for Kaon decays, $K \to \pi \phi (\to \text{inv)}$, \bmeson decays $B \to K \phi (\to \text{inv)}$ are not competitive with direct searches for long-lived particles if the dark Higgs decays predominantly into SM particles.

\subsubsection*{\texorpdfstring{$B \to X_S\scalar(\to e^+e^-,\mu^+\mu^-, \pi^+\pi^-,K^+K^-)$}{B to XSphi to ee, mumu, pipi, or KK} at \babar}
\label{subsec:limits_collider_light_babar}
\babar at the asymmetric $e^+e^-$ collider PEP-II in the US has used an integrated luminosity of 404~\fbinv to perform a search for mediator decays into $e^+e^-$, $\mu^+\mu^-$, $\pi^+\pi^-$, and $K^+K^-$ for various lifetimes~\cite{BaBar:2015jvu}.
The results are presented as 90\,\%~CL limits on the product of branching fractions $\bf(B\to X_s\scalar)\times \bf(\scalar\to x^+x^-)$ for different $\scalar$ lifetimes between 1\,\cm and 100\,\cm.
In contrast to the \belletwo analysis (see above), \babar did not reconstruct any exclusive \bmeson final states but only reconstructed the displaced $\scalar$ candidate.
This leads to larger backgrounds, but also to a higher production rate.
The results have been reinterpreted as limits on the dark Higgs mixing $\sin\theta$ for the muon and pion final state only~\cite{Winkler:2018qyg, Filimonova:2019tuy}.

\subsubsection*{\texorpdfstring{$B \to K^{(*)}\scalar(\to \mu^+\mu^-)$}{B to Kphi to mumu} at \lhcb}
\label{subsec:limits_collider_light_lhcb}
Using $pp$ collisions at the LHC, the \lhcb experiment has performed two dedicated searches for scalar particles produced in $b\to s$ transitions in \bmeson decays~\cite{LHCb:2015nkv, LHCb:2016awg} using an integrated luminosity of 3~\fbinv.
The scalar is searched for as a narrow di-muon resonance as a function of lifetime and includes both prompt and displaced scalar decays $\scalar \to \mu^+\mu^-$.
For all masses above $m_{\scalar} \gtrsim 0.5\,\gev$, the search in the final state $B^+\to K^+\scalar$ is more sensitive than $B^0\to \kstarz \scalar$.
The limits obtained by \lhcb have been rederived in Ref.~\cite{Winkler:2018qyg} using different scalar decay rates.
The limits at 95\,\%~CL are constraining scalar masses $2m_{\mu} < \mscalar \lesssim 4.8\gev$.
Ref.~\cite{Winkler:2018qyg} has also recast a published spectrum of prompt $B^+\to K^+\mu\mu$ decays \cite{LHCb:2012juf} which is, however, significantly weaker than the dedicated searches.

\subsubsection*{\texorpdfstring{$K^{\pm}\to \pi^{\pm}\scalar (\to\mathrm{inv})$}{K to pi phi to invisible} at NA62}
\label{subsec:limits_collider_light_na62}
The fixed target experiment NA62 uses the CERN SPS beam to search for the very rare decay $K^{\pm}\to \pi^{\pm}\nu\bar{\nu}$. 
The extracted $400\,\gev$ proton beam is dumped on a 40\,\cm long beryllium rod to produce a secondary beam containing a small fraction of about 6\% of charged kaons.
Using data collected until 2018, NA62 has also searched directly for invisible scalar particles or long-lived scalar particles that decay outside of the detector.
The limits  at 90\,\%\,CL are constraining scalar masses $\mscalar < 250$\,\mev with masses around $m_{\pi^0}$ vetoed from the search~\cite{NA62:2021zjw}; the region around $m_{\pi^0}$ is included in a second search specifically targeting this mass region~\cite{NA62:2020pwi}.

\subsubsection*{\texorpdfstring{$K^{\pm}\to \pi^{\pm}\scalar (\to\mathrm{inv})$}{K to pi phi to invisible} at E949}
\label{subsec:limits_collider_light_e949}
The combined results of the experiments E787 and E949 of the main search for the SM decay $K^{\pm}\to \pi^{\pm}\nu\bar{\nu}$ at Brookhaven National Laboratory (BNL) have also been analyzed by the collaboration as a search for invisible \scalar decays~\cite{BNL-E949:2009dza}.
Those experiments used a stopped \kaon beam and identified events with a charged \pion while rejecting events with any additional activity in the detector.
For the case of an unstable scalar particle, the experiments assumed a 100\,\% efficiency for detecting and vetoing such decays if the \scalar decayed inside the outer radius of the barrel veto.
The sensitivity of E787 and E949 hence increases with the lifetime of the scalar.
The limits constrain scalar masses $0 < \mscalar < 250$\,\mev with masses around $m_{\pi^0}$ vetoed from the search.
The experiment has presented exclusions in the \mscalar--$\bf(B)$ plane for different lifetimes between 100\,ps, and infinity.
The corresponding exclusion in the \mscalar--$\sin \theta$ plane has been calculated in Ref.~\cite{Winkler:2018qyg} and constrains the mixing angle for \scalar masses $m_{\scalar} \lesssim 210\,\mev$ with an insensitive vetoed region around $m_{\pi^0}$.

\subsubsection*{\texorpdfstring{$K^{\pm}\to \pi^{\pm}\scalar(\to e^+e^-, \mu^+\mu^-)$}{K to pi phi to ee or mumu} and \texorpdfstring{$\klong \to \pizero \scalar(\to e^+e^-, \mu^+\mu^-)$}{K) to pi0 phi to ee or mumu} at PS191}
\label{subsec:limits_collider_light_ps191}
The fixed target experiment PS191 used 19.2\,\gev protons from the CERN PS to search for heavy neutrinos\,\cite{Bernardi:1985ny, Bernardi:1987ek}. 
The experiment operated in the early 1980s and collected a dataset of about $8.6\times10^{18}$~\pot. 
Protons hitting the 80\,\cm long beryllium target produced mostly \pions and \kaons, that subsequently decayed in an approximately 50\,\m long, helium-filled decay volume, followed by a 5\,\m iron absorber. 
The detector was placed at a distance of about 128\,\m from the beryllium target, about 2.3\,$^{\circ}$ off-axis. The distance between the iron absorber and the detector was filled with dirt and soil. 
The detector consisted of eight $(6\times3)$\,\msq flash counter tracking planes perpendicular to the beam axis equally spaced over 12\,m, interspaced with helium bags. 
The tracking system was followed by a 7.2~radiation length deep calorimeter. 
The detector was triggered by a coincidence of the PS extraction signal and a hodoscope embedded into the calorimeter.
The trigger was efficient for pairs of electrons or muons.
Published limits on heavy neutrino decays were recast in Ref.~\cite{Gorbunov:2021ccu} as limits on a dark scalar produced in lepton-flavor violating $s\to d$ transitions in $K$-meson decays taking into account $\klong$ and $K^+$ production in the beryllium target, the decay volume, the surrounding dirt, and the iron absorber.
The limits at 90\,\%\,CL constrain the mass range $2m_e < m_{\scalar} < 250\,\mev$.
The recast is based on simulations to calculate signal yields including an approximate trigger simulation.
It conservatively ignores the presence of a focusing magnet for positively charged \kaons that would increase the scalar yield.

\subsubsection*{{\texorpdfstring{$K^+\to \pi^+ \scalar (\to e^+e^-,\mu^+\mu^-)$}{K to pi phi to ee, mumu} at \microboone}}
\label{subsec:limits_collider_light_mub}
The \microboone experiment is a liquid Argon time-projection chamber primarily designed for neutrino scattering measurements in the Booster Neutrino Beam (BNB) at Fermilab.
However, to search for a dark scalar the experiment used a dataset with an exposure of $1.93\times 10^{20}$\,\pot using the Fermilab NuMI neutrino beam~\cite{MicroBooNE:2021usw}.
The NuMi beam is produced by 120\,\gev protons hitting a graphite target producing secondary hadrons that decay in a helium-filled volume downstream of the target.
The remaining hadrons are then stopped in a 5\,\m deep hadron absorber, including charged \kaons which will decay at rest and produce dark scalars in flavor-changing $s\to d$ transitions.
\microboone then searches for visible decays of long-lived dark scalars.
The hadron absorber is located at a distance of 100\,\m and an angle of 125\,$^{\circ}$ with respect to the BNB direction, such that any particles coming from the absorber enter the detector in almost the opposite direction than neutrinos from the on-axis BNB.
With one observed candidate and a background expectation of about 2 events, they set 95\,\% CL upper limits for long lifetimes in the \mscalar--$\sin \theta$ plane.
Using the published model-independent limits on the product of branching ratio and lifetime, we derive 95\,\% CL upper limits for short lifetimes where \microboone loses acceptance to close the exclusion contour towards large values of $\sin \theta$.
The limits constrain scalar masses for $3\,\mev \lesssim m_{\scalar} \lesssim 210\,\mev$.
Using a similar experimental setup, \microboone has also searched for scalar decays into a pair of muons with an exposure of $7.01\times 10^{20}$\,\pot \cite{MicroBooNE:2022ctm} and placed 95\,\% CL upper limits.
These limits constrain scalar masses in the range $212\,\mev \lesssim m_{\scalar} \lesssim 275\,\mev$.

\subsubsection*{\texorpdfstring{$B\to X_s\scalar(\to e^+e^-, \mu^+\mu^-)$}{B to Ks phi} and \texorpdfstring{$K\to\pi\scalar(\to e^+e^-, \mu^+\mu^-)$}{K to pi phi} at CHARM}
\label{subsec:limits_collider_light_charm}
The CERN SPS beam has been used to directly search for visible axion-like particle decays into $e^+e^-$ or $\mu^+\mu^-$ in an empty decay region of 35\,m length and 9\,m$^2$ cross-section~\cite{CHARM:1985anb}.
This region was located parallel to the neutrino beam line used at the CHARM detector, about 10\,mrad off-axis.
The distance from the target to the detector was 480\,m.
The 400\,\gev proton beam was dumped on a copper target to produce mostly \kaons and a very small fraction of \bmesons.
The experiment used a dataset of about $2.4\times10^{18}$~\pot and did not observe any events with an efficiency of about 0.5.
The corresponding exclusion in the \mscalar--$\sin \theta$ plane has been calculated in Ref.~\cite{Winkler:2018qyg} taking into account \kaon absorption in the copper target\footnote{Recent studies in the context of future heavy neutral lepton searches at SHiP~\cite{GORBUNOV2020135817} suggest that a detailed simulation of Kaon interactions in the beam dump targets may yield a significantly smaller geometric acceptance and hence weaker limits than that assumed in ~\cite{Winkler:2018qyg}.}, but  neglecting \kaon regeneration by secondary interactions and assuming that the number of kaons escaping the target is negligible. 
The limits constrain the mixing angle for dark Higgs boson masses $m_{\scalar} \lesssim 280\,\mev$. 

\subsubsection*{$\klong \to \pizero \scalar(\to \mu^+\mu^-)$ at KTeV}
\label{subsec:limits_collider_light_ktev}
With the main scientific goal of measuring the branching fraction $\bf(\klong \to \pizero \mu^+\mu^-)$, the KTeV experiment used the Tevatron proton beam at Fermilab to search for two photons from a \pizero decay and two oppositely charged tracks, all coming from the same vertex.
The 800\,\gev proton beam was directed on a 30\,\cm BeO target.
A 65\,\m long instrumented vacuum decay volume was located 94\,\m downstream of the target.
KTeV observed two events consistent with the SM background expectation and placed an upper limit of \hbox{$\bf(\klong \to \pizero \mu^+\mu^-) < 3.8\times 10^{-10}$\,(90\,\% CL)}~\cite{KTEV:2000ngj}.
Since KTeV did not perform an invariant mass selection for the muon pair, the result can be reinterpreted as an upper bound on a promptly decaying dark Higgs boson using an estimated vertex resolution of 4\,\mm \cite{Dolan:2014ska}.
The corresponding exclusion in the \mscalar--$\sin \theta$ plane for $2 m_{\mu} < m_{\scalar} \lesssim 350\,\mev$ has been calculated in \cite{Winkler:2018qyg}.

\subsubsection*{\texorpdfstring{$\klong \to \pizero \scalar (\to\mathrm{inv})$}{KL to pi0 phi to invisible} at KOTO}
\label{subsec:limits_collider_light_koto}
The KOTO experiment at J-PARC is using a beam of \klong-mesons decaying in flight  to search for the rare decay \hbox{$\klong \to \pizero \nu\bar{\nu}$}.
The \klong beam is produced using 30\,\gev protons from the J-PARC main ring.
Protons hitting a gold target produce about $1.2\times10^{-7}$ \klong per proton on target.
The experiment used a dataset of about $3.05\times10^{19}$~\pot.
After an early analysis had revealed an anomalous excess of events~\cite{Shinohara:2020brf}, also called the `KOTO-anomaly', the final re-analysis yielded a result consistent with the background expectation. 
The corresponding upper limit is \hbox{$\bf(\klong\to\pi^0\nu\bar{\nu}) < 4.9\times 10^{-9}$\,(90\,\% CL)}~\cite{KOTO:2020prk}.
This result can be used to set limits on scalars that escape the KOTO detector before decaying into visible SM particles.
We  derive an exclusion region in the \mscalar--$\sin \theta$ plane for $m_{\scalar} < 2 m_{\mu}$ following the description in  Ref.~\cite{Egana-Ugrinovic:2019wzj} and using an average scalar momentum of 1.5\,GeV and a conservative efficiency estimate of $\varepsilon=0.75$ for all scalar masses to correct for the difference between the SM 3-body kinematics \hbox{$\klong \to \pizero \nu\bar{\nu}$} and the BSM 2-body kinematics \hbox{$\klong \to \pizero \scalar (\to\mathrm{inv})$}~\cite{KOTO:2018dsc}.

\subsection{Hints for dark Higgs bosons in hyperon decays}

In analogy to the rare meson decays discussed above, dark Higgs bosons can also be produced in hyperon decays, in particular $\Sigma^+ \to p \scalar$, which is sensitive to the flavor-changing effective coupling $h_{sd}$. In 2005, HyperCP searched for this decay in the $p \mu^+ \mu^-$ final state and observed three candidate events with an invariant di-muon mass just above the di-muon threshold~\cite{HyperCP:2005mvo}. These events are consistent with a two-body decay $\Sigma^+ \to p + X$ with $m_X \approx 214\,\mathrm{MeV}$ and $\text{BR}(\Sigma^+ \to p + X) \approx 3 \times 10^{-8}$.

While a translation of this result to fundamental model parameters is complicated by the uncertain hadronic form factors, it was quickly pointed out that the interpretation in terms of a light scalar Higgs boson was incompatible with constraints from rare meson decays, while an interpretation in terms of a light pseudoscalar remained viable~\cite{He:2006fr,He:2006uu,He:2005we,Geng:2005ra,Deshpande:2005mb}. In 2017, however, LHCb repeated the same measurement~\cite{LHCb:2017rdd}, finding a much broader distribution of di-muon invariant masses at a level consistent with the SM prediction. No evidence for a new particle produced in Hyperon decays was found, and the upper bound clearly excludes the HyperCP anomaly. 

\subsection{Future projections and proposed experiments}
\label{subsec:proposed_experiments}

\subsubsection*{Prospects for direct medium and high-mass dark Higgs searches}
\label{subsec:future_above_gev}

With $3 \abinv$ of $14 \tev$ $pp$ data collected at the future HL-LHC, the sensitivity to both medium and high-mass dark Higgs boson can be significantly improved. In the high mass range, the HL-LHC will allow improving the sensitivity by about a factor of three, probing mixing angles $\sin \theta$ e.g. down to 0.06 at $400 \gev$ and 0.17 at $1 \tev$~\cite{CidVidal:2018eel}. Mixing angles even reaching down to 0.01 for dark Higgs mass below $3 \tev$ could be tested with $20 \abinv$ of a future $14 \tev$ muon collider~\cite{Buttazzo:2018qqp}.  

In the intermediate mass range ($10 \gev \lesssim m_h \lesssim 100 \gev$), the HL-LHC will be able to probe mixing angles around 0.005~\cite{Carena:2022yvx}. A future electron-positron collider will provide further improvements on $\sin \theta$ using both associate production $e^+ e^- \to Z \phi$~\cite{Drechsel:2018mgd} and exotic decays of the SM-like Higgs boson $h \to \phi \phi$~\cite{Frugiuele:2018coc}. The former will be particularly important in the otherwise difficult-to-constrain mass range above $m_h / 2$.

\subsubsection*{Future experiments probing low-mass dark scalars}
\label{subsec:future_below_gev}
Many experiments have been proposed to improve the sensitivity to dark Higgs bosons in the GeV range. 
Among the most mature and promising proposals are SHiP~\cite{Aberle:2839677}, SHADOWS~\cite{Alviggi:2839484}, HIKE~\cite{CortinaGil:2839661}, FASER~2~\cite{Feng:2017vli}, DUNE~\cite{Berryman:2019dme}, \codexb~\cite{Aielli:2019ivi}, MATHUSLA~\cite{MATHUSLA:2020uve}, FACET~\cite{Cerci:2021nlb} and DarkQuest~\cite{Batell:2020vqn}. 
But also existing experiments such as \belletwo~\cite{Kachanovich:2020yhi,Filimonova:2019tuy} and LHCb~\cite{Craik:2022riw} promise substantial gains in sensitivity in coming years. 
A detailed review of all of these ideas is beyond the scope of this review, and we therefore refer to the activities of the Feebly Interacting Particle Physics center of the Physics Beyond Colliders Initiative at CERN\footnote{See Ref.~\cite{Beacham:2019nyx,Agrawal:2021dbo,Antel:2023hkf} and \url{https://pbc.web.cern.ch/fpc-mandate}.} instead.

\FloatBarrier
\section{Searches for decays into dark photons}
\label{sec:Higgs_dark_photon}
\begin{figure}%
\centering
\includegraphics[height=210px]{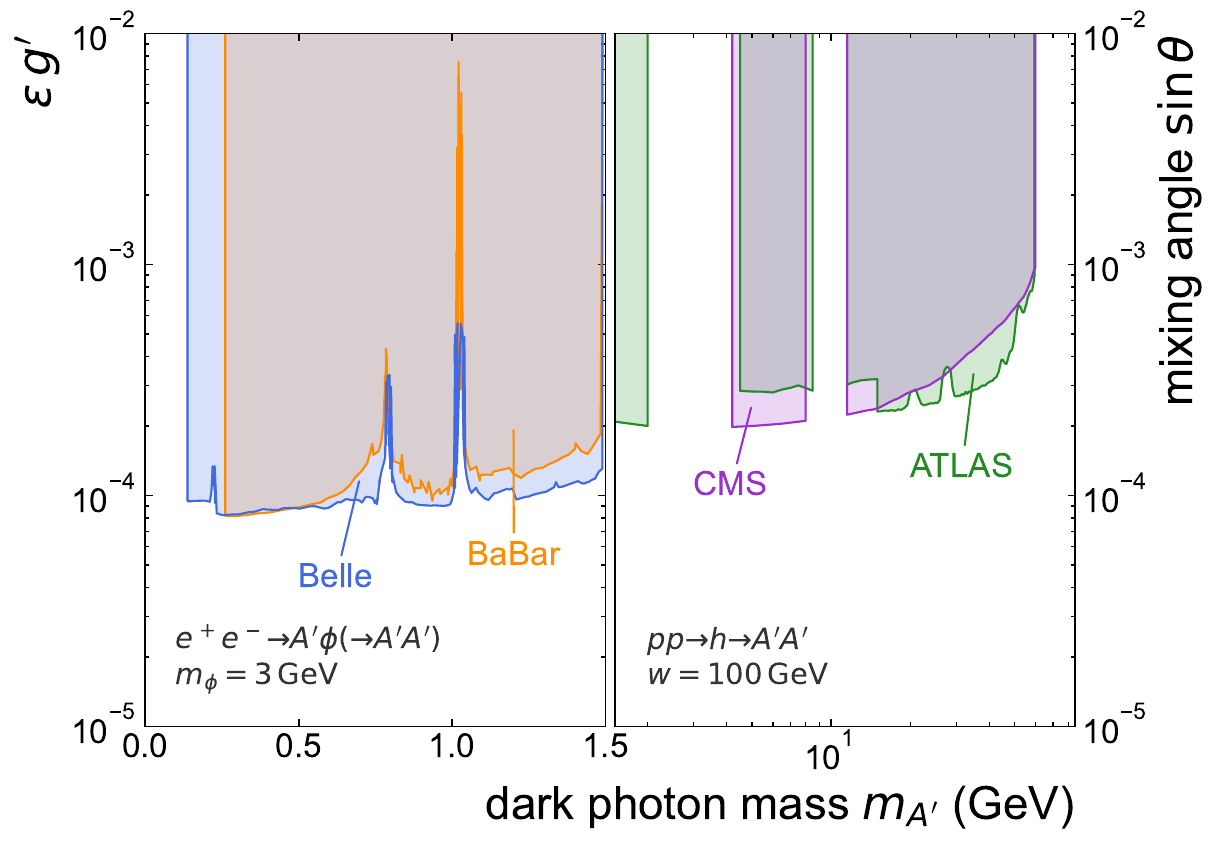}
\caption{\label{fig:DP} Constraints on dark Higgs models arising from searches for dark photons. The upper limits from \belleone~(90\,\% CL\,\cite{Jaegle:2015fme}) and \babar~(90\,\% CL\,\cite{BaBar:2012bkw}) in the left half of the plot result from searches for $e^+ e^- \to A' \phi (\to A' A')$, which constrains the product of the kinetic mixing parameter $\epsilon$ and the dark gauge coupling $g'$. 
The upper limits from CMS~(95\,\% CL\,\cite{CMS:2021pcy}) and ATLAS~(95\,\% CL\,\cite{ATLAS:2021ldb}) in the right half of the plot result from searches for $p p \to h \to A' A'$, which constrains the Higgs mixing parameter $\sin \theta$.}
\end{figure}

While it is plausible to assume that the dark Higgs boson is the lightest state in the dark sector, it is also possible to have $m_{A'} < m_\phi / 2$. 
This case has two important phenomenological consequences, as illustrated in figure~\ref{subfig:production_feynman:darkphoton_double}. 
First, the dark photon may give a relevant contribution to dark Higgs boson production via dark Higgs-strahlung. 
And second, both dark and SM-like Higgs bosons can now decay into pairs of dark photons, which would subsequently decay into SM particles (typically pairs of charged leptons or mesons) through mixing with the SM photon.\footnote{For a detailed discussion of the dark photon branching ratios, we refer to Ref.~\cite{Ilten:2018crw}.} Several experiments have searched for the resulting signatures. A summary of these searches is shown in figure~\ref{fig:DP}.

\subsection{Decays of the dark Higgs boson}
\label{subsec:dark_higgs_dark_photon}

\subsubsection*{$e^+e^-\to A'\phi, \phi\to A'A'$ at \belleone}
\belleone at the asymmetric $e^+e^-$ collider KEKB in Japan has used the full dataset of 977\,\fbinv at center-of-mass energies corresponding to the $\Upsilon(1S)$ to $\Upsilon(5S)$ resonances to search for dark Higgs-strahlung in fully visible final states $A'\to e^+e^-, \mu^+\mu^-, \pi^+\pi^-$ assuming $\mscalar>2m_{A'}$ with both promptly decaying dark Higgs and dark photons~\cite{Jaegle:2015fme}.
The process involves the coupling of the dark photon to SM particles via kinematic mixing $\epsilon$ with the SM photon, and the coupling of the dark photon to the dark Higgs $\alpha_\mathrm{D} \equiv g'^2 / (4\pi)$.
For dark photon masses $0.1\,\gev < m_{A'} < 3.5\,\gev$ and dark Higgs masses  $0.2\,\gev < \mscalar < 10.5\,\gev$, \belleone has set 90\%\,CL upper limits on the product coupling 
$\epsilon^2 \alpha_\mathrm{D}$ of the level of $10^{-10}$--$10^{-8}$. 
The Belle limits are the strongest limit to date in these mass ranges.

\subsubsection*{$e^+e^-\to A'\phi, \phi\to A'A'$ at \babar}
\babar has used a dataset of 516\,\fbinv at center-of-mass energies corresponding to the $\Upsilon(2S)$ to $\Upsilon(4S)$ resonances to search for dark Higgs-strahlung with a technique comparable to the aforementioned \belleone analysis~\cite{BaBar:2012bkw}.
For dark photon masses $0.25\,\gev < m_{A'} < 3.0\,\gev$ and dark Higgs masses  $0.8\,\gev < m_{A'} < 10.0\,\gev$, \babar has set 90\% confidence level upper limits on the product coupling 
$\epsilon^2 \alpha_\mathrm{D}$ of the level of $10^{-9}$--$10^{-8}$.

\subsection{Decays of the SM-like Higgs boson}

If the dark Higgs boson mixes with the SM-like Higgs boson, the latter inherits the coupling of the former to dark photons. 
For $m_{A'} < m_h/2$ it then becomes possible for SM-like Higgs bosons to decay into pairs of dark photons~\cite{Curtin:2014cca}. 
According to the Goldstone equivalence theorem, for $m_\phi, m_{A'} \ll m_h$ the corresponding decay width $\Gamma_{h \to A' A'}$ is equal to $\Gamma_{h \to \phi \phi}$ as given in eq.~\eqref{eq:Gammaphiphi}~\cite{Foguel:2022unm}. 
In other words, constraints on $\Gamma_{h \to A' A'}$ can be directly translated into bounds on the Higgs mixing parameter $\sin \theta$ (for a given value of the dark vev $w$). 

\subsubsection*{$pp \to h \to A' A'$ at CMS}

Using an integrated luminosity of $137 \fbinv$ of $13 \tev$ $pp$ collisions collected between 2016 and 2018, CMS has carried out a search for SM-like Higgs bosons decaying into a pair of dark photons, which subsequently decay into either $4\mu$, 2e2$\mu$, or 4e~\cite{CMS:2021pcy}. The analysis covers the dark photon mass range from $4.2 \gev$ to $60 \gev$ (except for a veto region around the $\Upsilon$ mass) and achieves a sensitivity of approximately $\sin \theta \lesssim 10^{-4}$--$10^{-3}$, depending on the assumed value of the dark photon mass but independent of the dark Higgs boson mass as long as $m_\phi \ll m_h$.

\subsubsection*{$pp \to h \to A' A'$ at ATLAS}

The corresponding ATLAS search~\cite{ATLAS:2021ldb} extends the mass range considered by CMS by an additional window between $1 \gev $ and $2 \gev$. While the search in the high mass range between 15 and $60 \gev$ explores $4\mu$, 2e2$\mu$, or 4e final states, the search in the low mass region up to $15 \gev$ focuses on $4\mu$ final states only. A sensitivity comparable to CMS is achieved.

\subsection{Future projections and proposed experiments}

The ATLAS and CMS bounds on exotic Higgs decays discussed above are expected to improve by an order of magnitude with HL-LHC data~\cite{Carena:2022yvx}, implying corresponding improvements for the bounds on $\sin \theta$~\cite{Liu:2017lpo}. At future electron-positron colliders, the most promising search strategy is analogous to the \belleone and \babar searches discussed above, but with the dark photon replaced by a $Z$ boson, i.e.\ $e^+ e^- \to Z + \phi (\to A' A')$. 

Finally, we would like to mention that it may also be possible to search for dark Higgs bosons decaying into dark photons in exotic $Z$ boson decays. Indeed, Ref.~\cite{Blinov:2017dtk} has pointed out that dark photons also mix with the SM $Z$ boson, and hence one can search for the exotic decay $Z \to A' \phi (\to A' A')$. The authors estimate that for dark Higgs and dark photon masses above the $B$ meson threshold, LHC searches for this decay mode may be competitive with direct bounds on the kinetic mixing parameter $\epsilon$ from LHCb and CMS.

\FloatBarrier
 \section{Searches for invisible dark Higgs decays}
\label{sec:invisible_Higgs}
\begin{figure}[t]
\centering
\includegraphics[height=210px]{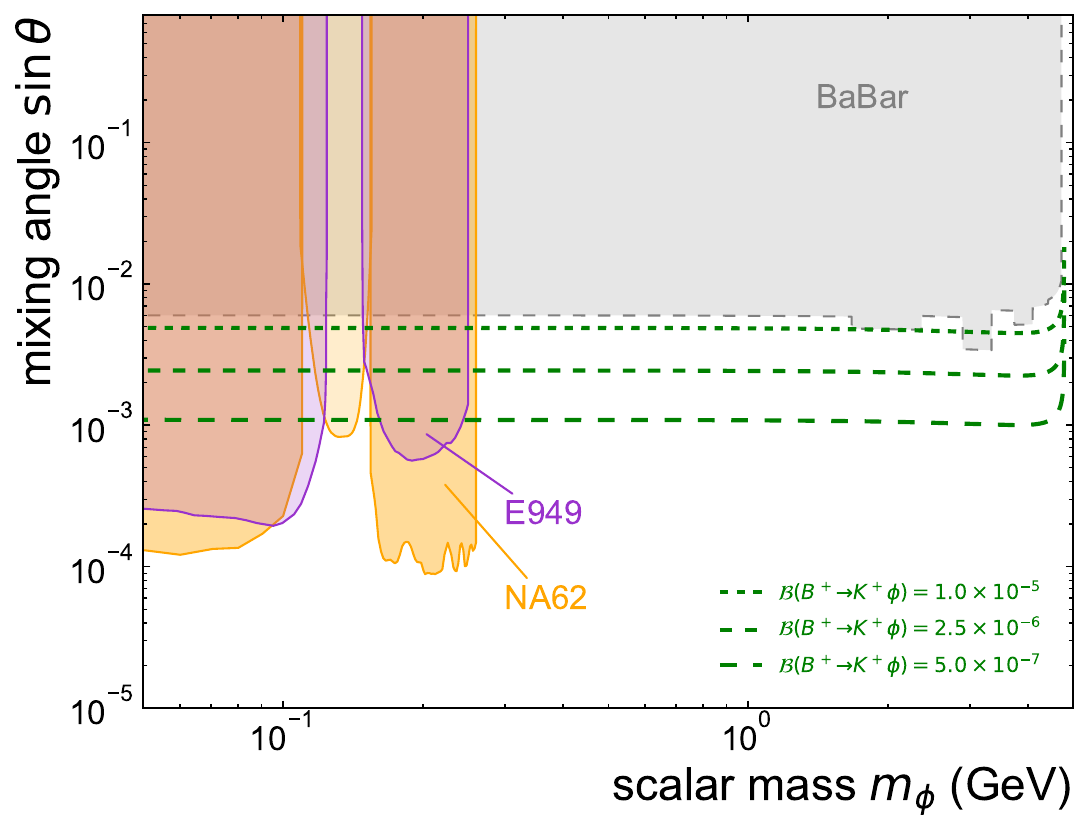}%
\caption{\label{fig:direct_higgs_0_5_invisible}  Upper  limits on the mixing angle $\theta$ as function of scalar mass $m_\phi$ from
NA62\,(95\,\%~CL\,\cite{Gorbunov:2021ccu}),
E949\,(90\,\%~CL\,\cite{BNL-E949:2009dza}), and
BaBar\,(95\,\%~CL\,\cite{BaBar:2013npw} and this work).
Constraints colored in gray with a dashed outline are reinterpretations not performed by the experimental collaborations and without access to raw data. The green dashed lines indicate the parameter combinations that correspond to branching ratios between $5 \times 10^{-7} \leq \mathcal{B}(B^+ \to K^+ \phi) \leq 1 \times 10^{-5}$.}
\end{figure}

If the dark Higgs boson couples to a DM particle with $m_\chi < m_\phi/2$, we expect invisible decays to dominate the branching ratios of the dark Higgs boson. While this would obviously make it much harder to conclusively discover dark Higgs bosons, we can exploit signatures with missing (transverse) energy to probe and constrain such models. The specific search strategies depend on the range of $m_\phi$.

\subsection{\texorpdfstring{$m_\phi > 10 \, \mathrm{GeV}$}{mphi greater than 10 GeV}}

For heavy dark Higgs bosons, the case of an invisibly decaying dark Higgs boson is experimentally very challenging, as it combines a moderate production cross-section with a rather unspecific final state. At the LHC, traditional DM searches for jets in association with missing energy are generally not sensitive to invisibly decaying dark Higgs bosons, because the production cross section is either suppressed by a small Yukawa coupling, if the dark Higgs is emitted from a light quark,  or by a loop factor, if the dark Higgs is emitted from a top-quark loop~\cite{Buckley:2014fba}. This makes it interesting to consider searches for heavy-quark final states in association with missing energy. 
Due to destructive interference effects in the production of a single top quark and a Higgs boson~\cite{Biswas:2012bd}, the strongest limits are provided by the $t\bar{t}+\text{MET}$ channel.

In the presence of mixing between the dark Higgs boson and the SM-like Higgs boson, the latter also obtains couplings to the DM particle proportional to $y_\chi^2 \sin^2 \theta$. As a result, both Higgs bosons may contribute to the $t\bar{t}+\text{MET}$ final state. While the relative magnitude of the two contributions depends on $y_\chi$, the general expectation is that the dark Higgs boson will give the dominant contribution for $m_\phi \ll m_h$ whereas the SM-like Higgs boson dominates the signal for $m_\phi \gg m_h$~\cite{Baek:2015lna}.\footnote{We note that for $m_\phi < m_h/2$ the SM-like Higgs boson may decay into a pair of dark Higgs bosons, giving an additional contribution to $t\bar{t} + \text{MET}$.} For $m_\phi \approx m_h$ both bosons give a relevant contribution, and the interference between them may be non-negligible~\cite{Ko:2018mew}.

\subsubsection*{$p p \to t \bar{t} \phi (\to \text{inv})$ at ATLAS}
ATLAS has performed a statistical combination~\cite{ATLAS:2022vpp} of final states with 0~\cite{ATLAS:2020dsf}, 1~\cite{ATLAS:2020xzu} or 2~\cite{ATLAS:2021hza} leptons each using an integrated luminosity of $139\,\mathrm{fb}^{-1}$. The result is an observed upper bound of $\sin\theta \lesssim 0.41$ for $m_\phi \ll m_h$. 

\subsubsection*{$p p \to t \bar{t} \phi (\to \text{inv})$ at CMS}
CMS has released a search for large missing transverse momentum in association with a $t \bar{t}$ pair using $137 \fbinv$ of data recorded at $\sqrt{s}=13 \tev$ between 2016 and 2018~\cite{CMS:2021eha}. The analysis combines previous searches in final states with 0~\cite{CMS:2021beq}, 1~\cite{CMS:2019ysk} or 2~\cite{CMS:2020pyk} leptons. 
While the primary target of the analyses is stop-quark pair production, the combined result is re-interpreted in a simplified DM model with scalar mediators and can be translated into limits on the mixing angle $\sin \theta \lesssim 0.39$ for $m_\phi \ll m_h$.

\bigskip

The result is thus comparable to that published by ATLAS and significantly weaker than the bound from the SM-like Higgs signal strength. We therefore do not show a figure for this case.

\subsection{\texorpdfstring{$m_\phi < 10 \, \mathrm{GeV}$}{mphi smaller than 10 GeV}}

For sufficiently light dark Higgs bosons, invisible decays lead to strong constraints from searches for $K \to \pi \phi (\to \text{inv)}$, which we already discussed in the context of long-lived dark Higgs bosons in section~\ref{sec:direct_Higgs}. In addition, it now becomes interesting to also consider rare \bmeson decays involving invisible final states. 

\subsubsection*{$B \to K \scalar (\to\mathrm{inv})$ at \babar}

In Ref.~\cite{BaBar:2013npw} \babar presents results in terms of $\mathcal{B}(B \to K \nu \bar{\nu})$ for different bins of $s_B = m_{\nu\nu}^2 / m_B^2$. 
Since the detector resolution for $s_B$ is at the level of a few percent, the signal from an invisibly decaying dark Higgs boson is expected to be strongly peaked at $s_B = m_\phi^2 / m_B^2$. 
We hence reinterpret the \babar result by assuming that the dark Higgs signal is fully contained in a single bin, multiplying the quoted uncertainties by 1.96 to obtain the approximate 95\%\,CL upper bound. 
We show the result of this reinterpretation together with the bounds from NA62 and E949 and the constraint from the SM-like Higgs signal strength in Fig.~\ref{fig:direct_higgs_0_5_invisible}. 

\subsubsection*{$e+ e^- \to A'(\to \mu^+ \mu^-) \phi (\to \text{inv})$ at KLOE-2}

Additional search strategies open up if dark Higgs bosons can be produced in association with a kinetically-mixed dark photon (see figure~\ref{subfig:production_feynman:darkphoton}) for dark Higgs boson masses as large as $0.5\gev$. In this case the production cross section becomes proportional to $\epsilon^2 \alpha_\mathrm{D}$, where $\alpha_\mathrm{D} = g'^2 / (4\pi)$ governs the probability for dark Higgs-strahlung. The final state is then a pair of SM fermions with invariant mass equal to the dark photon mass together with missing energy from the invisible dark Higgs boson decay. The first search for this signature was performed by the KLOE-2 experiment~\cite{KLOE-2:2015nli}, which used bins of $m_{\mu\mu}$ and $m_\text{miss}$ to reach a sensitivity of $\epsilon^2 \alpha_\mathrm{D} \lesssim 10^{-9} \text{--} 10^{-8}$ for sub-GeV dark photons. 

\subsubsection*{$e+ e^- \to A'(\to \mu^+ \mu^-) \phi (\to \text{inv})$ at \belletwo}
\belletwo has searched for dark Higgs bosons produced in association with a kinetically-mixed dark photon via $e^+e^-\to A'(\to\mu^+\mu^-)\phi(\to \mathrm{inv.})$ for dark Higgs boson masses as large as 4.5\gev using an early data set based on an integrated luminosity of $8.34\,\mathrm{fb}^{-1}$~\cite{Belle-II:2022jyy}. 
The best sensitivity ($\epsilon^2 \alpha_\mathrm{D} \lesssim 10^{-7}$) is achieved for heavy dark photons, where the energy in visible final states is maximized. 

\subsection{Future projections and proposed experiments}

At the moment, the LHC only probes models of invisibly decaying scalars that have an enhanced production cross-section compared to Higgs mixing. Nevertheless, in particular for the two-lepton final state, current searches are still limited by statistics, and improvements can be expected from HL-LHC~\cite{Haisch:2016gry}. Further improvements can be achieved with future lepton colliders, where the dark Higgs boson would be produced via Higgs mixing in association with a $Z$ boson. Ref.~\cite{Haghighat:2022qyh} showed that the ILC with $\sqrt{s} = 250 \, \mathrm{GeV}$ and an integrated luminosity of $2 \, \mathrm{ab}^{-1}$ may be sensitive to mixing angles as small as $\sin \theta \approx 0.02$ (see also Ref.~\cite{Ko:2016xwd}). Even stronger bounds may be obtained from future $Z$ factories, using the exotic decay $Z \to Z^\ast \phi$ followed by $Z^\ast \to \ell^+ \ell^-$ and an invisible dark Higgs boson decay~\cite{Liu:2017zdh}.

For dark Higgs boson masses below the \bmeson mass, significant advances are expected from \belletwo with larger data sets.
With an integrated luminosity of only 63\,\fbinv and using an inclusive tagging method, \belletwo already excluded $\bf(B^+\to K^+\nu\bar{\nu}) > 4.1\times 10^{-5}$ at the 90\% CL~\cite{Belle-II:2021rof}.
This inclusive method provides a higher signal efficiency and also sensitivity compared to the hadronic tag method used by \babar, but has almost no mass resolution.
As a result, this method is ideal to constrain the three-body decay $B^+\to K^+\nu\bar{\nu}$, where the missing mass of the neutrino pair follows a broad distribution, but is less sensitive than the hadronic tag method for the two-body decay $B^+\to K^+\phi$, where the narrow peak in the missing mass provides a key handle to distinguish signal and background (see Ref.~\cite{Ferber:2022rsf} for a detailed discussion in the context of a similar model). 
This also makes it difficult to interpret the recent evidence for the decay $B^+\to K^+\nu\bar{\nu}$ observed by Belle II~\cite{Belle-II:2023esi} as a signal of dark Higgs bosons~\cite{McKeen:2023uzo}.\footnote{As pointed out in Ref.~\cite{Bird:2006jd}, it is possible to obtain a broad distribution of missing masses from the decay $B \to K \chi \bar{\chi}$ via an off-shell dark Higgs boson. The branching ratio scales parametrically as $\sin^2 \theta y_\chi^2 / m_\phi^4$, which is unobservably small (given experimental bounds on $\sin \theta$) unless there is a second contribution to the mass of the DM particle such that $y_\chi \gg m_\chi / w$.}

The \belletwo sensitivity to measure the SM $\bf(B^+\to K^+\nu\bar{\nu})$ for an integrated luminosity of 50~\abinv using a hadronic tag method like \babar is about 11\%~\cite{Belle-II:2018jsg}.
Assuming that \belletwo observes an SM-like branching fraction $\bf(B^+\to K^+\nu\bar{\nu})=(5.67\pm0.38)\times10^{-6}$~\cite{Parrott:2022zte} with that precision, one can exclude dark Higgs mixing angles down to about $\sin \theta > 10^{-3}$ (compare Fig.~\ref{fig:direct_higgs_0_5_invisible}).
For the inclusive tag a similar sensitivity of about 11\% is expected~\cite{Belle-II:2022cgf}, while combining the different tags could yield a sensitivity as low as 8\%.
A combined analysis of all available \bmeson decays $B^0\to K^0_S\phi$, $B^0\to K^{*0}\phi$, and $B^+\to K^+\phi$ in an optimized search for this two-body decay instead of the SM three-body decay may provide even better sensitivity.


\FloatBarrier
\section{Searches for dark Higgs bosons in events with missing energy}
\label{sec:Higgs_MET}

In the previous section, we considered dark Higgs models with an additional dark fermion $\chi$ with $m_\chi < m_\phi/2$. If this inequality is not satisfied, the dark Higgs boson cannot decay invisibly. Nevertheless, the dark fermions may still play an important role in the search for dark Higgs bosons if both are produced together. This can happen for example if a dark photon is produced at a collider and radiates off a dark Higgs boson before decaying into a pair of dark fermions (see figures~\ref{subfig:production_feynman:darkphoton} and~\ref{subfig:production_feynman:darkphoton_double}), or if a dark Higgs boson is radiated. Such events can be searched for at the LHC by looking for large missing transverse momentum produced in association with a dark Higgs boson.

As pointed out in Ref.~\cite{Duerr:2017uap}, for sufficiently heavy dark photon masses, the cross-section for $p p \to \chi \chi \phi$ may be comparable to the one for $p p \to \chi \chi j$, where the jet arises from initial state radiation. However, the decays of the dark Higgs boson lead to much more distinctive experimental signatures, such as boosted topologies, meaning that backgrounds can be more easily suppressed and greater sensitivities can be achieved. While in principle any decay mode of the dark Higgs boson may be of interest, only two been investigated so far: the decay of the dark Higgs boson into two $b$-quark jets as well as the decay into two vector bosons.

\subsubsection*{$pp \to \chi \bar{\chi} \phi (\to b \bar{b})$ at ATLAS}

Decays into bottom quarks dominate if the dark Higgs boson has a mass below about $135\,\mathrm{GeV}$. 
This decay mode was targeted by ATLAS in Ref.~\cite{ATL-PHYS-PUB-2019-032} by reinterpreting a search for DM produced in association with an SM-like Higgs boson based on an integrated luminosity of $79.8\,\mathrm{fb}^{-1}$~\cite{ATLAS:2018bvd}. The reinterpretation was made using the RECAST framework~\cite{Cranmer:2010hk} and considers different $E^\text{miss}_T$ bins, which correspond to either two small-radius $b$ jets with radius parameter $R = 0.4$ (resolved case) or a single large-radius jet with $R = 1.0$ (merged fat jet). Discrimination between signal and background is achieved by considering the invariant mass of the jet pair (or the fat jet), which for the signal is strongly peaked at $m_\phi$. Very recently, the same strategy was applied in Ref.~\cite{ATL-PHYS-PUB-2022-045} to a DM search based on the full 13~\tev $pp$ LHC dataset~\cite{ATLAS:2021shl} using active learning. This approach makes it possible to obtain approximate limits for values of the dark photon couplings different from the ones usually considered. The sensitivity of this re-interpretation exceeds that of Ref.~\cite{ATL-PHYS-PUB-2019-032} and is thus the only one shown in Fig.~\ref{fig:MET} for the $E_{T}^{\rm miss} + bb$ final state. 

\subsubsection*{$pp \to \chi \bar{\chi} \phi (\to VV \to q\bar{q}q\bar{q})$ at ATLAS}

If the dark Higgs boson decays dominantly into gauge bosons, a number of different search strategies become available. In Ref.~\cite{ATLAS:2020fgc} ATLAS considers the fully hadronic final state using $139\,\mathrm{fb}^{-1}$ of data, which offers the possibility to reconstruct the dark Higgs boson mass, in particular if the dark Higgs is produced with sufficient boost that all its decay products are merged into a so-called track-assisted reclustering (TAR) jet. Nevertheless, this final state suffers from large backgrounds from $V + \text{jets}$, which limit the achievable sensitivity.

\subsubsection*{$pp \to \chi \bar{\chi} \phi (\to WW \to q\bar{q}\ell \nu)$ at ATLAS}

Substantially higher sensitivity can be achieved in the semi-leptonic final state, where backgrounds are reduced at the expense of losing information about the dark Higgs boson mass. In Ref.~\cite{ATLAS:2022bzt}, ATLAS performs such a search both for the case that the two jets are resolved or that they are merged into a TAR jet. Special care is taken for highly boosted Higgs bosons where the lepton potentially merges with the TAR jet so that only hadronic objects are included and the reconstructed mass close to the $W$ boson is preserved. They construct a kinematic variable $m_s^\text{min}$, which satisfies $m_s^\text{min} < m_\phi$ and can be used to separate signal and background.

\subsubsection*{$pp \to \chi \bar{\chi} \phi (\to WW \to q\bar{q}\ell \nu)$ and $pp \to \chi \bar{\chi} \phi (\to WW \to \ell \nu \ell \nu)$ at CMS}

In Ref.~\cite{CMS:2023cdc} CMS considered the same final state for the full run-2 dataset using a BDT to separate signal from background. 
In addition, the analysis also considers the fully leptonic final state, for which a kinematic variable called $m_T^{\ell_\text{min},p_T^\text{miss}}$ can be constructed, which peaks at the $W$ boson mass for the background but takes larger values for the signal.

\begin{figure}[t]%
\centering
\includegraphics[height=210px]{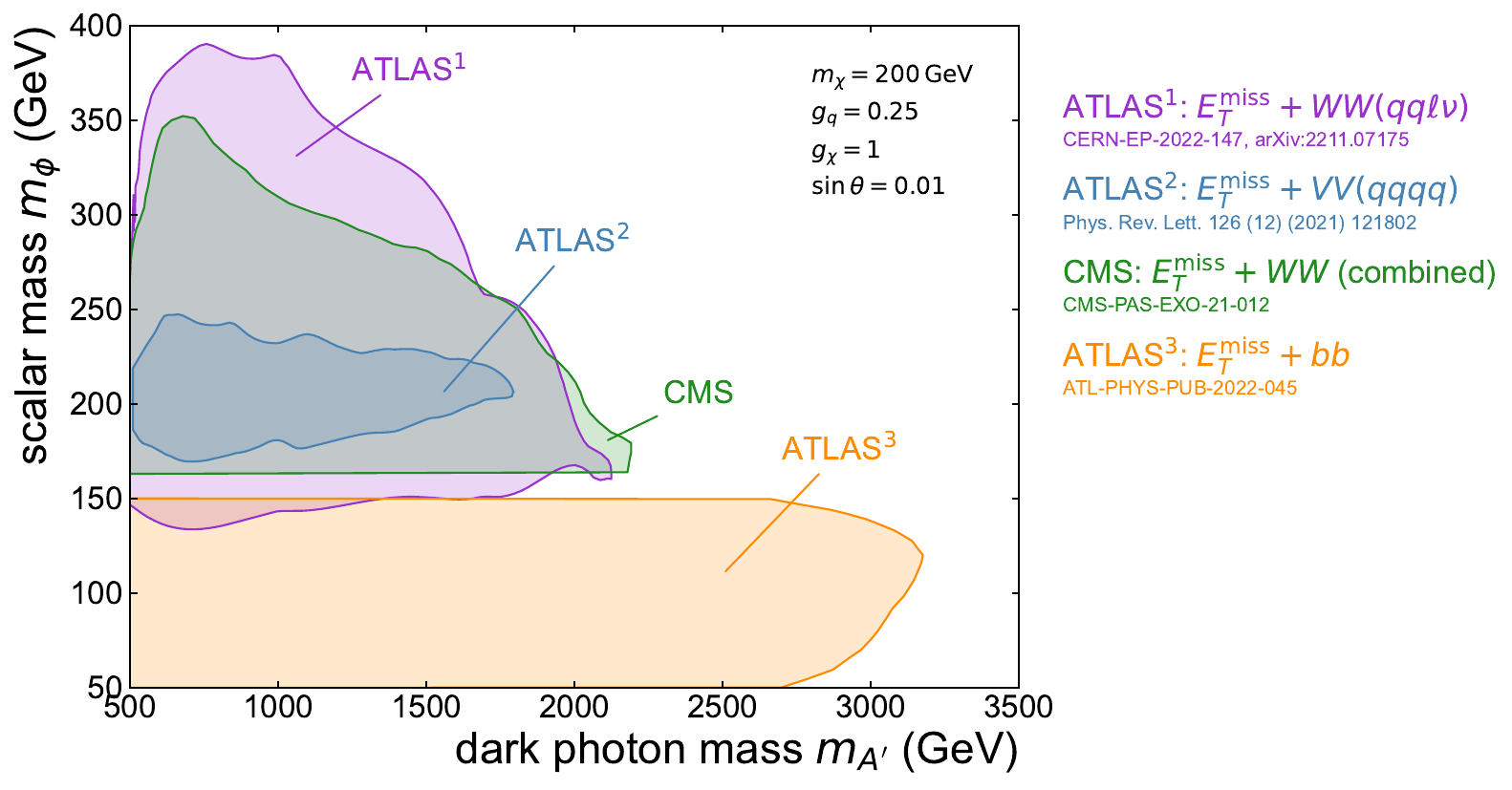}
\caption{\label{fig:MET} \hbox{95\,\% CL} exclusion limits on a visibly decaying dark Higgs boson produced in association with an invisibly decaying dark photon $A'$ (or $Z'$) as a function of the two masses $m_{A'}$ and $m_\phi$ from ATLAS~\cite{ATL-PHYS-PUB-2022-045, ATLAS:2020fgc, ATLAS:2022bzt} and CMS~\cite{CMS:2023cdc}. Here we assume that SM quarks carry direct charge $g_q = 0.25$ under the new $U(1)'$ gauge group, while the DM particles  carry charge $g_\chi = 1$ and have a mass $m_\chi = 200\,\mathrm{GeV}$ such that the decay $\phi \to \chi \bar{\chi}$ is kinematically forbidden. The Higgs mixing angle has been fixed to $\sin \theta = 0.01$, but its precise value is inconsequential.}
\end{figure}

\bigskip

A comparison of existing results from the LHC is shown in figure~\ref{fig:MET}. In this plot the dimensionality of the parameter space has been reduced by fixing $m_\chi = 200 \, \mathrm{GeV}$ and $\sin \theta = 0.01$, with the precise value of the Higgs mixing angle being inconsequential as long as it is large enough to ensure prompt dark Higgs boson decays. Furthermore, it is assumed that the dark photon has flavor-universal direct couplings to SM quarks given by $g_q = 0.25$ and a direct coupling to the dark fermion $g_\chi = 1$. For a detailed discussion of how these choices affect the exclusion limits, we refer to Ref.~\cite{Duerr:2017uap}.

\FloatBarrier
\section{Proposed searches for dark Higgs bosons in events with displaced vertices}
\label{sec:Higgs_DV}
In this section we discuss future searches for dark Higgs bosons in events with displaced vertices, arising from the decay of a neutral long-lived particle.\footnote{We emphasize that neutral long-lived particles may also decay exclusively into neutral final states, such as photons or neutral pions, which would not allow for the reconstruction of the decay vertex. These decays are much more difficult to identify and distinguish from background, and will therefore not be considered further in this review.} 
Such a signature may arise from processes analogous to the ones considered in section~\ref{sec:Higgs_MET}, if one of the particles produced in association with the dark Higgs boson is long-lived rather than stable. 
This set-up is realized in models of Pseudo-Dirac dark matter, where two states $\chi_{1,2}$ with mass splitting $\delta$ couple off-diagonally to the dark photon, leading to the decay chain $A^\prime \to \chi_1 \chi_2 (\to \chi_1 + f \bar{f})$, where $f \bar{f}$ denotes a pair of charged SM states produced via an off-shell dark photon. 

Since $\chi_2$ decays into a three-body final state, the invariant mass of the visible final states follows a broad distribution, which makes it difficult to distinguish signal from background even if a displaced vertex can be reconstructed. 
Collider searches for Pseudo-Dirac dark matter are therefore most promising if additional particles can be produced in association with the $\chi_1$--$\chi_2$ pair. 
Such additional particles may either come from initial-state radiation~\cite{Duerr:2019dmv,CMS:2023ojs} or if the dark photon emits dark Higgs-strahlung before decaying~\cite{Duerr:2020muu}. 
In the latter case, we can expect another pair of charged SM states, emitted either from a prompt or displaced vertex (see the left diagram in Fig~\ref{subfig:production_feynman:idmdh}). 
This second pair has an invariant mass equal to $m_\phi$, leading to a striking signature that can be readily distinguished from backgrounds. Ref.~\cite{Duerr:2020muu} studied the sensitivity of \belletwo to this signature, finding promising prospects already for early data sets. 

Depending on the mass spectrum of the dark sector, there can be a number of variations of the signature discussed above. 
For example, if $m_\phi < \delta$ and if the dark Higgs boson is responsible for generating the mass splitting, it may also be possible to produce a dark Higgs boson in the decay $\chi_2 \to \chi_1 + \phi$~\cite{Darme:2018jmx}. 
If the excited dark matter state is long-lived, this decay chain leads to a displaced vertex even if the dark Higgs decays promptly (see the diagram in Fig~\ref{subfig:production_feynman:idm}). 
In contrast to the case discussed above, the invariant mass of the particles originating from the displaced vertex would reconstruct to the dark Higgs boson mass. 
Conversely, if $m_\phi > 2 m_{\chi_2}$, it may be possible to produce long-lived particles in the decays of a dark Higgs boson, produced either via Higgs mixing or dark Higgs-strahlung. 
This process has been studied in Ref.~\cite{Li:2021rzt} in order to estimate the sensitivity of \faser to inelastic dark matter. 
It is however not possible in this set-up to reconstruct the properties of the dark Higgs boson from the visible decay products.

To conclude this section, let us mention a search for long-lived dark photons produced in dark Higgs boson decays carried out by the \atlas experiment~\cite{ATLAS:2019tkk} with an integrated luminosity of $36.1\,\mathrm{fb}^{-1}$. 
In this search, the dark photons are assumed to be several orders of magnitude lighter than the dark Higgs boson, such that they are produced with high boost and their decay products resemble a displaced jet. 
\atlas then searches for pairs of such displaced dark photon jets.
While the primary target is exotic decays of the SM-like Higgs boson, the search is also interpreted in terms of heavier scalar resonances.

\FloatBarrier
\section{Constraints from cosmology, astrophysics and non-collider experiments}
\label{sec:cosmology_and_astro}
To conclude our discussion of dark Higgs bosons, we discuss the role that they may play in the early universe, in astrophysical systems, and as mediators of the interactions of dark matter.

\subsection{Cosmological and indirect detection constraints}
\label{sec:cosmo_and_indirect}
The first question of interest is whether dark Higgs bosons would thermalize with the plasma of SM particles. While thermalization can in principle happen already in the unbroken phase (i.e.\ before the two Higgs bosons acquire their vacuum expectation values), it is most efficient once the two Higgs bosons can mix with each other. The various processes that contribute to the thermalization of dark Higgs bosons have been studied in detail in Ref.~\cite{Evans:2017kti}. The dominant contributions are found to come from $h \to \phi\phi$ (if kinematically allowed) and $q g \to q \phi$. Broadly speaking, the conclusion is that for $\sin \theta \gtrsim 10^{-7}$ the dark Higgs boson will enter into thermal equilibrium with the SM thermal bath, although this equilibrium may not be maintained at low temperatures when the number densities of the SM Higgs bosons and heavy quarks become Boltzmann suppressed.

If dark Higgs bosons enter into thermal equilibrium with the SM thermal bath, it is essential that they decay or annihilate away before the beginning of Big Bang Nucleosynthesis (BBN) in order to avoid spoiling the successes of standard cosmology. A commonly quoted~\cite{Krnjaic:2015mbs} requirement on the lifetime of the dark Higgs boson is $\tau < 1\,\mathrm{s}$, but closer analysis reveals some dependence of this bound on the dark Higgs mass and the decay products. For example, for a dark Higgs boson mass of $100\,\mathrm{MeV}$, the bound is $\tau < 0.2\,\mathrm{s}$~\cite{Depta:2020zbh}, while even stronger bounds are expected if pions are produced in the dark Higgs boson decays~\cite{Boyarsky:2020dzc}.

If, on the other hand, the mixing angle is so small that the dark Higgs bosons do not enter into thermal equilibrium, it may still be possible to produce them non-thermally via the freeze-in mechanism~\cite{Chu:2011be,Berger:2016vxi}. In this case, the constraints from BBN are relaxed considerably, and it is in fact possible for dark Higgs bosons to be stable on cosmological scales, such that they may constitute the dominant form of dark matter~\cite{Heeba:2018wtf,Fradette:2018hhl,Mondino:2020lsc}. The leading constraints on this scenario stem from searches for mono-energetic x-ray and $\gamma$-ray lines produced in the decays of dark Higgs bosons~\cite{Heeba:2018wtf}, bounds on dark matter self-interactions from the Bullet Cluster~\cite{Markevitch:2003at} and, for dark Higgs boson masses in the keV range, from warm dark matter bounds~\cite{DEramo:2020gpr,Decant:2021mhj}. The viable mixing angles are so small that the dark Higgs boson would be unobservable in laboratory experiments, and therefore we do not consider this case further.\footnote{We note that it is also possible for dark Higgs bosons to decay between BBN and recombination, provided their abundance is small enough, or for the dark Higgs boson to decay into DM particles. For more detailed discussions, we refer to Refs.~\cite{Ibe:2021fed,Ghosh:2022fws}.}

Even if the dark Higgs boson itself does not constitute a dark matter candidate, it may help to explain the observed dark matter relic abundance  via the freeze-out mechanism either as a mediator for dark matter annihilations (i.e.\ processes of the form $\text{DM DM} \to \phi^{(\ast)} \to \text{SM SM}$~\cite{Binder:2022pmf}) or as final state in the secluded annihilation process $\text{DM DM} \to \phi \phi$~\cite{Bondarenko:2019vrb}. The latter case is particularly predictive, as it fixes the coupling $y_\chi$ between dark matter and dark Higgs bosons as a function of the dark matter mass (with negligible dependence on $m_\phi$ for $m_\phi \ll m_\text{DM}$). To first approximation, the observed relic abundance is reproduced for $y_\chi \approx 0.04 \times \sqrt{m_\text{DM} / \text{GeV}}$~\cite{Kahlhoefer:2017umn}.\footnote{We emphasize that this estimate assumes that the dark sector remains in thermal equilibrium with the SM thermal bath during freeze-out. While this assumption is questionable if the dark Higgs is the only mediator between the dark and the visible sector~\cite{Bringmann:2020mgx}, it is generally plausible in models that also contain a dark photon.}

If annihilations into dark Higgs bosons are responsible for setting the dark matter relic abundance, these processes may still be observable in the present universe through indirect detection experiments. Indeed, it has been shown in Refs.~\cite{Ko:2014loa,Ko:2014gha,Ko:2015ioa} that this set-up can in principle fit the $\gamma$-ray Galactic Centre Excess~\cite{Daylan:2014rsa,Fermi-LAT:2017opo} for dark matter masses around $100\,\mathrm{GeV}$. However, in the standard set-up with a fermionic dark matter particle, annihilation into a pair of Higgs bosons is a $p$-wave process, meaning that the cross-section today is suppressed proportional to the dark matter velocity squared, $v_\text{DM}^2 \sim 10^{-6}$. Ref.~\cite{Bell:2016fqf} pointed out a promising alternative, namely the $s$-wave process $\text{DM} + \text{DM} \to \phi + Z'$. Indeed, if this process is kinematically allowed, one finds strong constraints from Fermi-LAT observations of dwarf spheroidal galaxies~\cite{Fermi-LAT:2015att}. Furthermore, Ref.~\cite{An:2016kie} showed that bound-state formation may lift the $p$-wave suppression and lead to strong indirect detection constraints (see also Ref.~\cite{Ko:2019wxq}). More complex indirect detection signatures have been discussed in Ref.~\cite{Jodlowski:2021xye}.

An entirely different avenue to probe dark Higgs bosons is to search for the stochastic gravitational wave background emitted during the phase transition that leads to spontaneous symmetry breaking~\cite{Athron:2023xlk}, which is found to be first-order in parts of the parameter space~\cite{Addazi:2017gpt,Hashino:2018zsi}. Whether or not the resulting signal may be observable with near-future gravitational wave observatories depends on both the strength of the transition as well as on the temperature of the dark sector relative to the SM thermal bath~\cite{Breitbach:2018ddu,Ertas:2021xeh}. It was shown in Ref.~\cite{} that in dark Higgs models the peak frequency of the gravitational wave signal is strongly correlated with the DM relic abundance. In addition, the phase transition may also lead to the production of dark matter particles through the decay of dark Higgs bosons crossing the bubble walls that separate the two phases, even if $m_\phi \ll m_\text{DM}$~\cite{Azatov:2021ifm}.

Finally, let us mention that, analogous to the idea of Higgs inflation~\cite{Bezrukov:2007ep}, dark Higgs bosons may play the role of the inflaton~\cite{Lebedev:2011aq,Kim:2014kok}. Compared to Higgs inflation, these models are less sensitive to the precise values of the SM-like Higgs boson and top-quark masses, but still predict values of the spectral index and the tensor-to-scalar ratio compatible with constraints from Planck~\cite{Planck:2018jri}.

\begin{figure}[t]%
\centering
\includegraphics[width=0.75\textwidth]{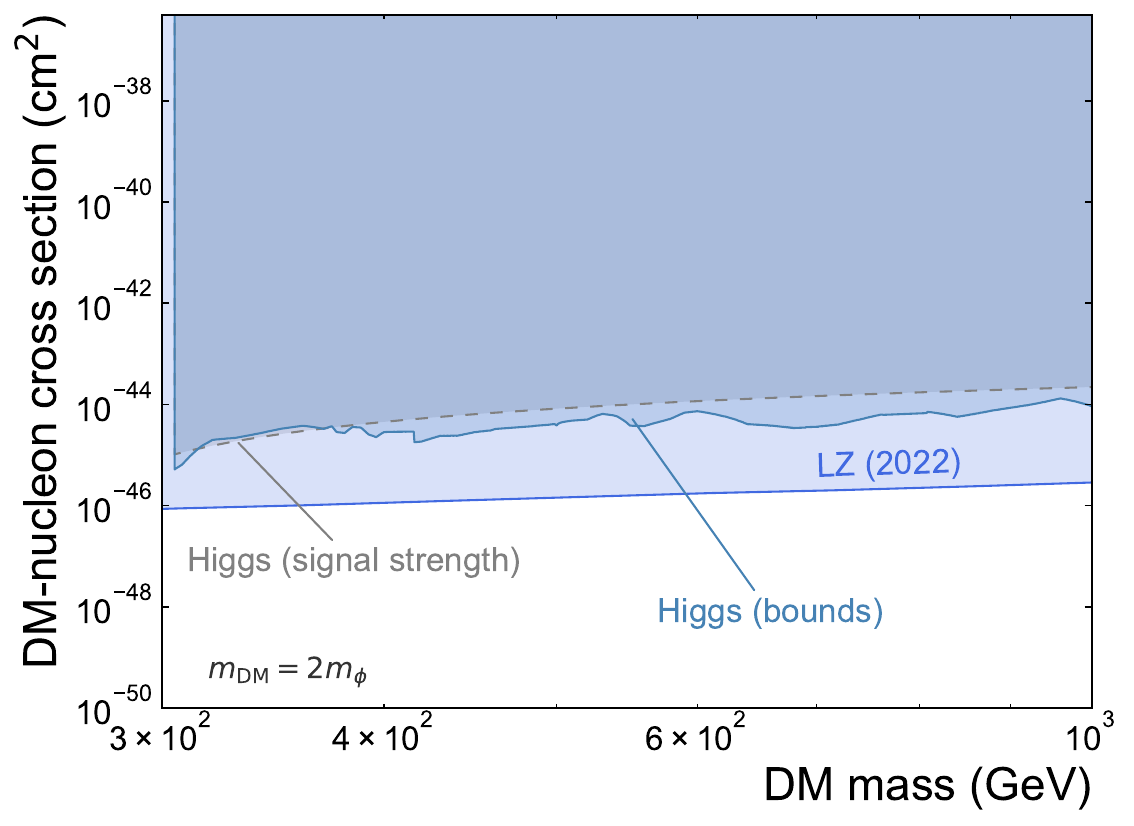}
\caption{\label{fig:DD_high} Upper  limits on the spin-independent DM-nucleon cross-section $\sigma$ as function of DM mass $m_{\mathrm{DM}}$ for
heavy dark Higgs bosons from LZ~\cite{LZ:2022ufs}.
For limits from direct dark Higgs searches see Sec.\,\ref{sec:direct_Higgs}, and for limits from Higgs signal strength see Sec.~\ref{sec:Signal_strength}. 
Constraints colored in gray with a dashed outline are reinterpretations not performed by the experimental collaborations and without access to raw data.}
\end{figure}

\begin{figure}[t]%
\centering
\includegraphics[width=0.75\textwidth]{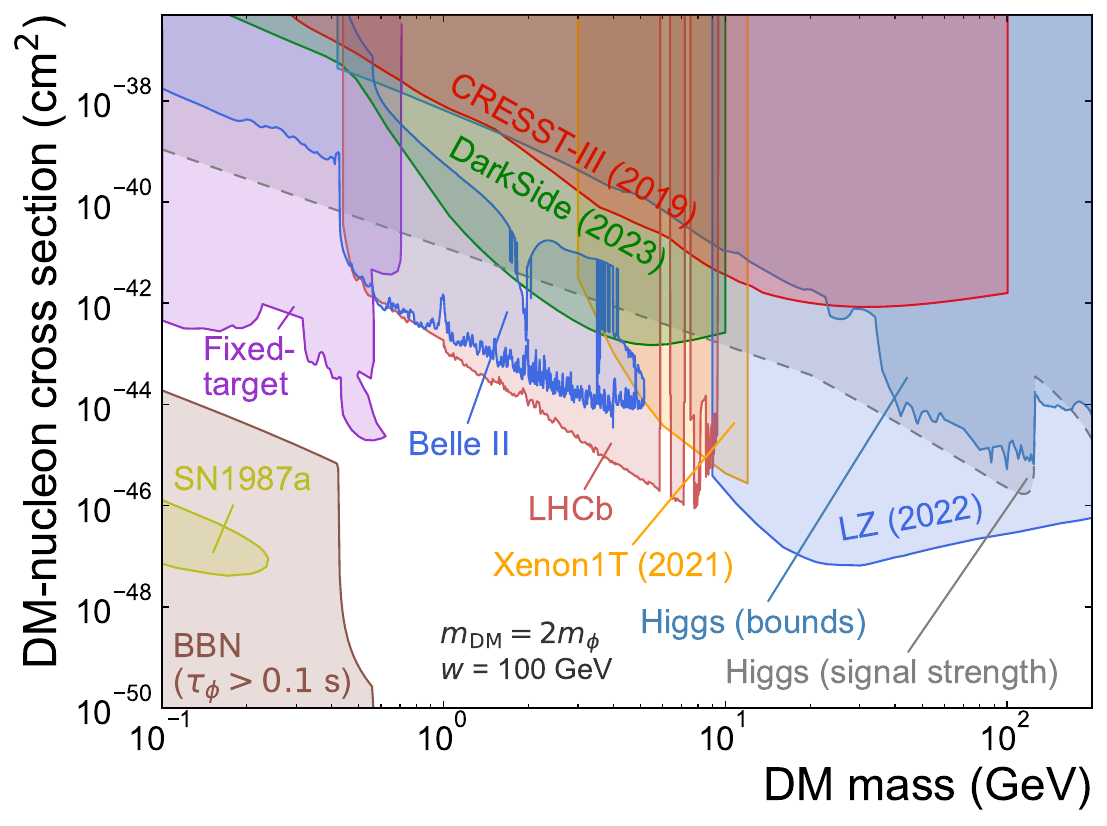}
\caption{\label{fig:DD_low} Upper  limits on the spin-independent DM-nucleon cross-section $\sigma$ as function of DM mass $m_{\mathrm{DM}}$ for
light dark Higgs bosons from \cresstiii~\cite{CRESST:2019jnq}, DarkSide~\cite{DarkSide:2022dhx}, XENON1T~\cite{XENON:2020gfr}, and LZ~\cite{LZ:2022ufs}.
For limits from direct dark Higgs searches, limits from flavor experiments and fixed-target experiments see Sec.\,\ref{sec:direct_Higgs}, and for limits from Higgs signal strength see Sec.~\ref{sec:Signal_strength}. We also indicate the approximate lower bound on $\sin \theta$ imposed by BBN~\cite{Depta:2020zbh} and the parameter region excluded by the duration of the neutrino signal from SN1987a~\cite{Dev:2020eam}.
Constraints colored in gray with a dashed outline are reinterpretations not performed by the experimental collaborations and without access to raw data.}
\end{figure}

\subsection{Astrophysical and direct detection constraints}
\label{sec:astro_and_dd}
Sub-GeV dark Higgs bosons can be produced in astrophysical systems and thereby constitute an exotic cooling mechanism~\cite{Raffelt:1996wa}. Of particular interest in this context is SN1987a, which is commonly interpreted as a core-collapse supernova explosion. Dark Higgs bosons can be produced in the hot and dense supernova core via nucleon-nucleon bremsstrahlung ($N N \to N N + \phi$)~\cite{Krnjaic:2015mbs}. If the dark Higgs bosons escape from the supernova core without decaying or being absorbed, they would reduce the luminosity in neutrinos, in conflict with observations~\cite{Kamiokande-II:1987idp}. A careful recent analysis~\cite{Dev:2020eam} showed that this consideration excludes mixing angles in the range $\sin \theta \sim 10^{-6}\text{--}10^{-5}$ for $m_\phi \lesssim 100\,\mathrm{MeV}$.
For sub-MeV dark Higgs bosons, even stronger bounds arise from the cooling of horizontal branch stars and red giants, which require $\sin \theta \lesssim 10^{-9}$ for $m_\phi < 10\,\mathrm{keV}$~\cite{Hardy:2016kme}.

Furthermore, dark Higgs bosons may mediate the scattering of dark matter particles on nuclei, leading to potentially observable signals in direct detection experiments.\footnote{Due to Higgs mixing, DM-nucleon scattering can also be mediated by the SM Higgs boson. In the limit $m_\phi \gg m_h$ one recovers so-called Higgs portal DM models, which we will not discuss further here. Instead, we refer to Ref.~\cite{GAMBIT:2018eea} for a recent analysis of these models.} The DM-nucleon scattering cross-section at zero momentum transfer is given by~\cite{Baek:2011aa}
\begin{equation}
\sigma_N = \frac{\mu_N^2 \, f_N^2 \, m_p^2 \, y_\chi^2 \, \cos^2 \theta \, \sin^2 \theta}{2\pi \, v^2}\left(\frac{1}{m_\phi^2} - \frac{1}{m_h^2}\right)^2 \; ,
\end{equation}
where $\mu_N$ is the reduced DM-nucleon mass and $f_N\approx 0.3$ is the effective Higgs-proton coupling~\cite{Bishara:2017nnn} with some uncertainty due to the strange-quark content of the nucleon~\cite{FlavourLatticeAveragingGroup:2019iem}.

For $m_\phi \gg m_\text{DM}$ the DM annihilation cross section is also proportional to $y_\chi^2 \cos^2 \theta \sin^2\theta / m_\phi^4$, such that direct detection constraints can be directly compared to the relic density requirement $\Omega_\text{DM} \approx 0.12$~\cite{Planck:2018vyg}. The outcome of this comparison is that, given the very strong constraints from liquid-xenon-based experiments like XENONnT~\cite{XENON:2023sxq}, PandaX-4T~\cite{PandaX-4T:2021bab} and LZ~\cite{LZ:2022ufs}, it is very difficult to reproduce the observed relic density unless the nuclear recoil energy is below threshold, corresponding to sub-GeV DM masses. For such small DM masses, however, the relic density requirement is typically incompatible with constraints on the invisible Higgs decay $h \to \chi \chi$, see Ref.~\cite{GAMBIT:2018eea}.

The more attractive option is therefore that $m_\phi < m_\text{DM}$, and the relic density requirement simply fixes $y_\chi$ as a function of $m_\text{DM}$ (see above). 
In this case, direct detection constraints place an upper bound on the mixing angle $\theta$ for given values of $m_\text{DM}$ and $m_\phi$. 
These bounds get stronger as the ratio $m_\text{DM} / m_\phi$ increases. 
In the following, when showing direct detection constraints, we will conservatively take $m_\text{DM} / m_\phi = 2$.\footnote{We note that this choice is generally compatible with the assumption that the DM particle obtains its mass from the dark Higgs boson via a suitable choice of the dark vev $w$.} In figures~\ref{fig:DD_high} and ~\ref{fig:DD_low} we show the comparison of direct detection constraints with other constraints on dark Higgs bosons. 
The leading constraints at 90\% confidence level stem from \cresstiii~\cite{CRESST:2019jnq}, DarkSide~\cite{DarkSide:2022dhx}, XENON1T~\cite{XENON:2020gfr}, and LZ~\cite{LZ:2022ufs}.
These constraints are found to be stronger than the ones from invisible Higgs decays. 

No relevant constraints on $\sin \theta$ arise from searches for DM-electron scattering due to the smallness of the corresponding Yukawa coupling. However, these searches are very sensitive to the couplings $g'$ and $\epsilon$ of the dark photon, offering an alternative avenue to constrain dark Higgs boson models.

 \section{Outlook}
\label{sec:outlook}
Dark Higgs bosons arising from the spontaneous breaking of a $U(1)'$ gauge extension of the SM provide a simple and well-motivated framework to connect the key frontiers of modern particle physics:
\begin{itemize}
    \item detailed studies of the production and decay modes of the SM-like Higgs boson;
    \item LHC searches for new particles at the TeV scale, for example, extended Higgs sectors;
    \item precision measurements of rare meson decays;
    \item exotic collider signatures with high-multiplicity final states and/or missing energy;
    \item searches for long-lived particles at high-intensity facilities.
\end{itemize}
At the same time, dark Higgs bosons may be directly connected to the DM puzzle. They provide a viable mechanism to reproduce the observed DM relic abundance and predict detectable signals in laboratory experiments and astrophysical observations.

On all of these fronts, tremendous progress can be expected over the next two decades, as the LHC and Belle II continue to collect data and new experiments with high-intensity or ultra-low backgrounds will be constructed. Among the most exciting opportunities in the near future is the measurement of $B \to K + \text{invisible}$ at Belle II and the construction of a new beam-dump and/or a new forward-physics facility at CERN. 

At the same time, progress in the theoretical description of dark Higgs bosons is needed to match the experimental improvements. For example, there is still considerable uncertainty in the lifetime and branching ratios of dark Higgs bosons with mass at the GeV scale, and in the constraints imposed on dark Higgs bosons from stellar cooling and supernova explosions. Finally, the rapidly growing field of gravitational wave detection motivates more detailed studies of the phase transition that triggers spontaneous symmetry breaking. 

In this review, we have shown how results from many different experiments and measurements can be presented or reinterpreted in terms of dark Higgs bosons, both in minimal models, where the dark Higgs boson is the only kinematically accessible state and in simple extensions, where dark photons and DM particles may give rise to distinctive experimental signatures. This comparison makes the complementarity of different experimental strategies apparent and highlights the need to either agree on benchmark models across collaborations or to make all relevant data publicly available. We hope that the plots shown in this review will be populated with many more exclusion contours and, eventually, closed confidence regions.

\section*{Acknowledgements}

It is a great pleasure to thank (in alphabetical order) Thomas Biek{\"o}tter, Pawel Guzowski, Belina von Krosigk, Gaia Lanfranchi, Ulrich Nierste, Maksym Ovchynnikov and Martin W.~Winkler for discussions, and Thomas Biek{\"o}tter, Sascha Dreyer, Maksym Ovchynnikov, Kai Schmidt-Hoberg, and Susanne Westhoff for feedback and comments on earlier versions of the manuscript.
This work was supported by the Helmholtz-Gemeinschaft Deutscher Forschungszentren (HGF) through
the Young Investigators Group VH-NG-1303, and by
the Deutsche Forschungsgemeinschaft (DFG) through
the Emmy Noether Grant No. KA 4662/1-2 and grant
396021762 -- TRR 257. 
Moreover, the authors acknowledge funding by the DFG under Germany‘s Excellence Strategy – EXC 2121 “Quantum Universe” – 390833306.

\end{document}